\documentclass[qsecnum,amsmath,preprintnumbers,superscriptaddress,nofootinbib,aps,prd,10pt,a4paper]{revtex4-1}
\pdfoutput=1
\usepackage[utf8]{inputenc}
\usepackage[english]{babel}
\usepackage{amsmath}
\usepackage{amsfonts}
\usepackage{subcaption}
\usepackage{amssymb}
\usepackage{amsthm}
\usepackage{mathrsfs}
\usepackage{hyperref}
\usepackage{color}
\usepackage{tikz}
\usepackage{feynmp-auto}
\usepackage{bm}
\usepackage{graphicx}
\usetikzlibrary{matrix}
\usepackage{caption}

 \def\be{\begin{equation}}
\def\ee{\end{equation}}
 \def\ba{\begin{align}}
\def\ea{\end{align}}
\def\bea{\begin{eqnarray}}
\def\eea{\end{eqnarray}}

\newcommand{\di}{\mathrm d}
\def\a{\alpha}
\def\b{\beta}

\def\m{\mu}
\def\n{\nu}

\newcommand{\bseq}{\begin{subequations}}
\newcommand{\eseq}{\end{subequations}}

\begin{document}

\preprint{FTUAM-20-28} 
\preprint{IFT-UAM/CSIC-20-190}

\title{{\bf Hawking Radiation from Universal Horizons}}
\author{Mario Herrero-Valea}
\email[]{mherrero@sissa.it}
\address{SISSA, Via Bonomea 265, 34136 Trieste, Italy and INFN Sezione di Trieste}
\address{IFPU - Institute for Fundamental Physics of the Universe \\Via Beirut 2, 34014 Trieste, Italy}

\author{Stefano Liberati}
\email[]{liberati@sissa.it}
\address{SISSA, Via Bonomea 265, 34136 Trieste, Italy and INFN Sezione di Trieste}
\address{IFPU - Institute for Fundamental Physics of the Universe \\Via Beirut 2, 34014 Trieste, Italy}

\author{Raquel Santos-Garcia}
\email[]{raquel.santosg@uam.es}
\address{Departamento de F\'isica Te\'orica and Instituto de F\'isica Te\'orica, IFT-UAM/CSIC,
Universidad Aut\'onoma de Madrid, Ciudad Universitaria de Cantoblanco, 28049 Madrid, Spain}

\begin{abstract}

The persistence of a suitable notion of black hole thermodynamics in Lorentz breaking theories of gravity is not only a non-trivial consistency test for such theories, it is also an interesting investigation {\em per se}, as it might help us identifying the crucial features at the root of these surprising laws governing such purely gravitational objects. In past investigations, controversial findings were presented in this sense. With the aim of settling this issue, we present here two complementary derivations of Hawking radiation in geometries endowed with universal horizons: a novel feature of back holes in Lorentz breaking theories of gravity which reproduces several properties normally characterizing Killing horizons. We find that both the derivations agree on the fact that the Hawking temperature associated to these geometries is set by the generalized universal horizon peeling surface gravity, as required for consistency with extant derivations of the first law of thermodynamics for these black holes.  We shall also comment on the compatibility of our results with previous alternative derivations and on their significance for the survival of the generalized second law of black hole thermodynamics in Lorentz breaking theories of gravity.

\end{abstract}
%%%%%%%%%%%%%%%%%%%%%%%%%%%%%%%%
%%%%%%%%%%%%%%%%%%%%%%%%%%%%%%%%
\maketitle
%%%%%%%%%%%%%%%%%%%%%%%%%%%%%%%%
%%%%%%%%%%%%%%%%%%%%%%%%%%%%%%%%
\section{Introduction}

Local Lorentz invariance of space-time (LLI) is at the basis of our present understanding of Nature. It is a fundamental symmetry of the standard model of particle physics as well as a funding pillar of General Relativity. Nonetheless, the search for a definitive theory of quantum gravity has led in the last decades to a renewed questioning about this symmetry and its role in Nature. First of all, being the Lorentz group non-compact we cannot say we have tested it completely, and we also know that LLI is strictly related to ultraviolet divergences in quantum field theory, being the latter a direct consequence of the assumption that the spectrum of field degrees of freedom is boost invariant. Moreover, hints of possible departures from Lorentz invariance came by several quantum gravity scenarios and, albeit non-conclusive, further stimulated the study of the phenomenological consequences of these departures from standard physics (see e.g.~\cite{Mattingly:2005re,Liberati:2013xla} for extensive reviews).

Missing a definitive theory of quantum gravity, it has been a common approach to take a bottom up attitude and adopt an effective field theory approach to Lorentz violations by adding systematically Lorentz breaking operators to different sectors of the Standard Model and then use current observations to constrain them~\cite{Mattingly:2005re,Liberati:2013xla}. Lorentz breaking can be limited to the boost sector, in a Riemannian geometry, at the cost of introducing a preferred vector field (normally a unit, timelike one) that selects a preferred frame\footnote{Let us stress that is not strictly speaking necessary for a fundamental theory of quantum gravity to be endowed with a preferred frame for it to induce, below the Planck energy, an effective field theory with broken boost invariance. For this to happen it is sufficient that the UV and IR theories do not share the same Lorentz group, and this can happen for example if the space-time is akin to a condensate of fundamental constituents. See e.g.~\cite{Fagnocchi:2010sn} for a realization of this scenario within an analogue gravity framework and \cite{Oriti:2016acw} for a specific quantum gravity proposal entailing a ``space-time condensate" scenario.}.
This in turn, together with the desire to preserve background independence, led to the development of extensions of General Relativity which include, also in vacuum, the dynamics of an ``aether" field and not just of the metric. The most general quadratic action for a metric and the aether is the one of the so called Einstein-Aether theory of gravity~\cite{Jacobson:2000xp}. It is possible however to further require that such a vector selects as well a preferred foliation: this can be easily enforced by requiring it to be the gradient of a scalar and hence ``hyper-surface orthogonal" and vorticity free. The action, in this case, takes the form of the low energy action of Ho\v rava Gravity (which in principle entails not only mass dimension four operators but also several others, up to mass dimension eight, to guarantee power counting renormalizability).

It is worth stressing that while we do have fairly stringent constraints on deviations from Lorentz invariance in particle physics (see e.g.~\cite{Liberati:2013xla}), we do not have comparable constraints on the gravitational sector. Indeed, currently detected gravitational waves are too low energy to constrain in any way terms of mass dimensions larger than four, and even for the latter, they severely constrain only two of the three basic parameters of the theory~\cite{Gumrukcuoglu:2017ijh}. Fortunately, when dealing with a physical theory it is not only possible to test it by using observations and experiments but it is also possible to probe its self-consistency.

Black holes represent the purest gravitational objects in nature and their thermodynamic aspects are still nowadays considered one of the few phenomena allowing us to have a glimpse into the possible merging of the gravitational and quantum realms.  It is then not surprising that they can also play a crucial role in testing the self-consistency of Lorentz breaking theories of gravity. Furthermore, proving or disproving the persistence of black holes' thermodynamics in theories with Lorentz invariance would provide invaluable insight to what lies at the root of this tantalizing feature of gravity, and hence it would be relevant beyond the actual interest in such alternative theories with a preferred frame.

Eddington\footnote{The Nature of the Physical World (1915), chapter 4.} famously said that an unavoidable consistency test for a theory is to check whether the second law of thermodynamics holds true. This test seemed however miserably failed by the aforementioned Lorentz breaking theories of gravity when it was realized that violations of LLI seem to directly lead to violation of the Generalized Second Law (GSL) of black hole thermodynamics. The basic mechanism, in this case, is based on the fact that if different fields on a black hole space-time are endowed with different limit speeds, then in general they will experience different horizons and ergoregions. This mismatch can then be used to generate a violation of the second law, for example, by constructing a classical and quantum perpetuum mobile, or by devising a setup whose only outcome is to lower the total entropy of the universe~\cite{Dubovsky:2006vk,Eling:2007qd,Jacobson:2008yc,Blas:2011ni}. 

While this could have spelled doom for this whole class of theories, in a surprising twist it was soon recognized that black hole solutions in these frameworks have an internal structure that is quite different from the standard black holes of General Relativity.  First of all, it was shown that classical arguments for GSL violation can be circumvented once the dynamics of the non-relativistic theory (for gravity and matter) is properly taken into account~\cite{Benkel:2018abt}. Also, for what concerns Hawking radiation based arguments, it appeared soon clear that due to the possibility of non-relativistic dispersion relations for matter fields and gravitons, in such theories the Killing horizon does not capture the notion of the causal boundary as in standard black-hole space-times, as it can be realized by including the aforementioned higher order terms in the matter action (so to have not only different limit speeds but also superluminal signalling). Moreover, it was recognized that, static, spherically-symmetric black-hole solutions, in both Einstein-Aether and Ho\v rava theories, contain a special hypersurface that acts as a genuine causal boundary because it traps all possible signals, even those traveling at arbitrarily high speeds\footnote{Such a special hypersurface, a leaf of the preferred foliation and hence of constant khronon field, can be found also in slow rotating solutions in Ho\v rava gravity while no extension exists for rotating black holes in Einstein-Aether gravity where a priory no preferred foliation needs to exist~\cite{Barausse:2012qh,Barausse:2015frm}.}. This special hypersurface has been called the ``universal horizon”. Indeed it was also proven that the typical peeling structure shown by null rays at the Killing horizon in General Relativity is reproduced now at the universal horizon by the generalized causal rays moving under the influence of the metric {\em and} the khronon field with the peeling being characterized (albeit not completely fixed) by the surface gravity of the universal horizon, $\kappa_{\text {\sc uh}}$~\cite{Cropp:2013sea}. 

It is evident that such finding might play a crucial role in the restoration of the GSL in non-relativistic frameworks as the presence of a universal temperature associated to the UV completion of the framework considered in ~\cite{Dubovsky:2006vk,Eling:2007qd,Jacobson:2008yc,Blas:2011ni} would solve at the root any possibility of construction of a perpetuum mobile, at least based on Hawking radiation. Indeed, further support for a restoration of black hole thermodynamics via universal horizons was lent by the finding that a generalized first law of black-hole mechanics can be associated to universal horizons~\cite{Berglund:2012bu,Berglund:2012fk,Liberati:2017vse}, instead of the usual Killing horizon, and by requiring that the temperature associated to the universal horizon has to be fixed by their  surface gravity $\kappa_{\text {\sc uh}}$. Noticeably, such association was derived independently in~\cite{Berglund:2012bu} using the tunneling formalism of Ref.~\cite{Parikh:1999mf}, and later on generalized to arbitrary truncated power law dispersion relations, $\omega\sim p^N$ for large $p$, in \cite{Ding:2015fyx,Ding:2016srk,Cropp:2016gkn, Liberati:2017vse}.~\footnote{Note however, that the aforementioned thermodynamic interpretation seems to have a non-trivial dependence on the asymptotic structure of the solutions~\cite{Bhattacharyya:2014kta,Basu:2016vyz}, with problems in asymptotically anti-de Sitter space-times, and that there are still
open issues for rotating black holes \cite{Liberati:2017vse}.}

Despite this growing evidence, still many points have remained unclear. In particular, if the conjectured temperature associated with the universal horizon would be relevant for observers outside the black hole, or if instead the emitted radiation would be somehow reprocessed at the Killing horizon in an energy and species dependent way. In this context, the results reported in \cite{Michel:2015rsa} seemed to surprisingly contradict the accumulated evidence for a universal horizon thermodynamics, as they implied that the temperature associated to Ho\v rava black holes was, as usual, the one set by the peeling surface gravity of the Killing horizon (see e.g.~\cite{Cropp:2013zxi} for a detailed discussion of different notions of surface gravity).

An extra motivation to understand the thermal properties of black holes in Einstein-Aether and Ho\v rava Gravity comes from holography in Lifshitz systems through the AdS/CFT duality. These are systems similar to CFTs but which exhibit an anisotropic scaling between time and space. Initially, they were proposed to be described by bulk theories combining a Proca field and a dilaton field. The former condenses along a direction -- in a similar fashion to the Aether here -- thus breaking boost invariance and leaving a non-relativistic theory. However, it was later shown by \cite{Griffin:2012qx} that some properties related with conformal anomalies that were not easily captured by these duals were emerging naturally in Ho\v rava Gravity, opening thus the window to a Lifshitz/Lifshitz holography \cite{Cheyne:2017bis}. In this context, a complete understanding of thermality within the gravitational theory in the bulk could lead to important developments and applications for Lifshitz theories at finite temperature on the boundary.

In order to settle the issues described above, in this paper we shall rederive the Hawking temperature associated to spherically symmetric black holes in Ho\v rava gravity using two different methods. We shall show that this is indeed associated to the generalized $\kappa_{\rm peeling}$ as found in \cite{Ding:2015fyx,Ding:2016srk,Cropp:2016gkn}. Also, we shall show that the different outcome of the investigation reported in \cite{Michel:2015rsa} is not due to any technical issue but rather to a different choice of the vacuum state. 
%Which vacuum state choice is the right one, apart from consistency with other calculations, is something that will have to be eventually fixed by a detailed description of the formation of such black holes in a gravitational collapse.

The plan of the paper is the following. In chapter \ref{sec:LIVBH} we shall revise black holes in Lorentz violating theories. In chapter \ref{sec:HRColl} we shall review the usual derivation in a relativistic setting of Hawking radiation in a collapsing shell geometry. In section \ref{sec:HRUH} we shall apply the same method, by a suitable generalization of physical rays, to the case of a geometry endowed with a universal horizon. We shall show that in this case, the relevant temperature for Hawking radiation is the one set by the universal horizon. To investigate further the compatibility of this result with the investigation reported in \cite{Michel:2015rsa} we revisit their calculation in chapter~\ref{sec:WKB}, showing that their and our results can be reproduced with a suitable different choice of the vacuum state. Finally, we provide in chapter \ref{sec:theral} an extension of our results to more general dispersion relations, and we discuss their implications in the conclusions.

%%%%%%%%%%%%%%%%%%%%%%%%%%%%%%%%
%%%%%%%%%%%%%%%%%%%%%%%%%%%%%%%%

\section{Lorentz violating black holes}
\label{sec:LIVBH}
We will consider the following action describing Einstein-Aether gravity
\begin{align}\label{eq:action}
S=-\frac{1}{16\pi G}\int \di^4x\sqrt{|g|}\ \left(R+ c_1 \nabla_\m U_\n \nabla^\m U^\n +c_2 (\nabla_\m U^\m)^2 + c_3 \nabla_\m U^\n \nabla_\n U^\m +c_4 a^\m a_\m 
+\lambda (U_\m U^\m-1)\right),
\end{align}
where $a^\m=U^\n \nabla_\n U^\m$.

Here $G$ is the Newton's constant and $R$ the scalar curvature of the manifold. The couplings $c_i$ are dimensionless and parametrize the departures from General Relativity in terms of the dynamics of the aether vector $U^\m$, which is forced to have unit norm $U_\m U^\m =1$ by the Lagrange multiplier $\lambda$, thus breaking boost invariance. The coupling $c_2$ is constrained to be positive to ensure the absence of ghosts \cite{Blas:2009qj}. Here we are interested in static black hole solutions, which are only known analytically for aether fields that are hyper-surface orthogonal. In that case, $c_1$ can be always absorbed into redefinitions of the rest of the couplings, so we will fix $c_1=0$ from now on. Incidentally, in that case, this action corresponds to the low energy limit of Ho\v rava Gravity, known as \emph{khronometric theory}\footnote{In this case, the couplings are usually renamed as $c_2=\lambda$, $\alpha=c_4$, $\beta=c_3$.} \cite{Blas:2010hb}, in which the aether is described in terms of a single scalar degree of freedom, the khronon $T$ 
\begin{align}
U_\m=\frac{\partial_\m T}{\sqrt{\partial_\a T \partial^\a T}}.
\end{align}
Note that $U_\m U^\m=1$ is automatic from this and therefore solutions enjoy a preferred time direction given by the integral lines of $U^\m$. As a consequence, the theory enjoys a reduced symmetry group with respect to General Relativity. Instead of full \emph{Diffeomorphism} invariance, the dynamic is invariant only under those diffeomorphisms which preserve the global time direction, dubbed \emph{FDiff} \cite{Blas:2010hb}.

In \eqref{eq:action} we have set a possible cosmological constant to vanish. We do this for simplicity since we are interested in understanding Hawking radiation. Although formally this forbids the application of our results to contexts like AdS/CFT, we expect them to be easily translated to those cases. The role of a cosmological constant for the black hole solutions that we are going to study in the following would be only to regulate the large distance behavior of the metric and therefore, the dynamics of internal structures, such as horizons, whose characteristic length is typically much smaller than the curvature radius of a possible AdS embedding, should not be affected.

Regarding the viability of \eqref{eq:action} as a realistic theory of Gravity, it must be noted that although Lorentz violations are highly constrained in the matter sector of the Standard Model \cite{Liberati:2013xla}, current bounds on gravitational dynamics are still not as competitive. The allowed parameter space has been quickly reduced in recent years, but there is still room for departures to be accommodated. In particular, the couplings in \eqref{eq:action} are constrained to satisfy $|c_3|\lesssim 10^{-15}$ from the observation of multi-messenger signals from GW170817 \cite{Monitor:2017mdv,Gumrukcuoglu:2017ijh,Oost:2018tcv}, while Solar System tests \cite{Will:1993hxu,Will:2014kxa,Blas:2011zd,Bonetti:2015oda} demand $|c_4|\lesssim 10^{-7}$. An additional theoretical constrain, demanding that black holes moving slowly with respect to the aether remain regular except for their central singularity, requires $c_3=c_4=0$ exactly \cite{Ramos:2018oku}.  On the other hand, measurements of the abundance of elements during Big Bang Nucleosynthesis \cite{Carroll:2004ai,Yagi:2013ava,Yagi:2013qpa} bound $|c_2|\leq 0.1$.

Here, however, we will be looking to a different region of the parameter space, not in the physical region but advantageous for our purposes here, given by $c_4=2c_3=-2c_2$. For this particular value of the coupling constants, the action \eqref{eq:action} admits an analytic spherical static black hole solution given by the Schwarzschild metric
\begin{align}\label{eq:metric}
\di s^2=F(r)\di t^2 -\frac{\di r^2}{F(r)}-r^2 \di \Omega^2
\end{align}
with $\di \Omega^2=\di \theta^2+\sin^2 \theta \di \phi^2$ and 
\begin{align}
F(r)=1-\frac{2\mu}{r},
\end{align}
which will thus allow us to perform computations in an analytical manner, without having to resort to numerical methods, necessary for other values of the couplings. However, the main feature of the solution that leads to the phenomenon of interest for our work here is the presence of the universal horizon, which is present for generic values of all $c_i$. Thus, we expect our analytical findings to also hold in the whole parameter space, at least schematically. Finally, this is the same solution considered in \cite{Michel:2015rsa}, and thus it allows us to lay a direct comparison of our results and methods against those of them.

The metric \eqref{eq:metric} is complemented by an aether configuration given by
\begin{align}
U_\a \di x^a=\left(1-\frac{\mu}{r}\right)\di t+\frac{\mu }{r-2\mu} \di r,
\end{align}
and its orthogonal vector
\begin{align}
S_\a \di x^\a=-\frac{\mu}{r}\di t+\frac{\mu-r}{r-2\mu}\di r.
\end{align}

As previously advertised, this aether configuration is hyper-surface orthogonal and it is normed to unity, while its orthogonal vector satisfies $S^\m S_\m=-1$. 
Note that there is a sign arbitrariness in choosing $S^\m$ which we fix by demanding it to approach $S^\m=\delta^\m_r$ for large radii. The metric has a Killing horizon at $r_{\text {\sc kh}}=2\mu $ for particles moving at light speed ($c^2=1$) and a time-like Killing vector given by
\begin{align}
\chi_\a \di x^\a =\frac{r-2\mu}{r} \di t,
\end{align}
as well as the usual curvature singularity when $r=0$.

\begin{figure}
  \includegraphics[scale=.20]{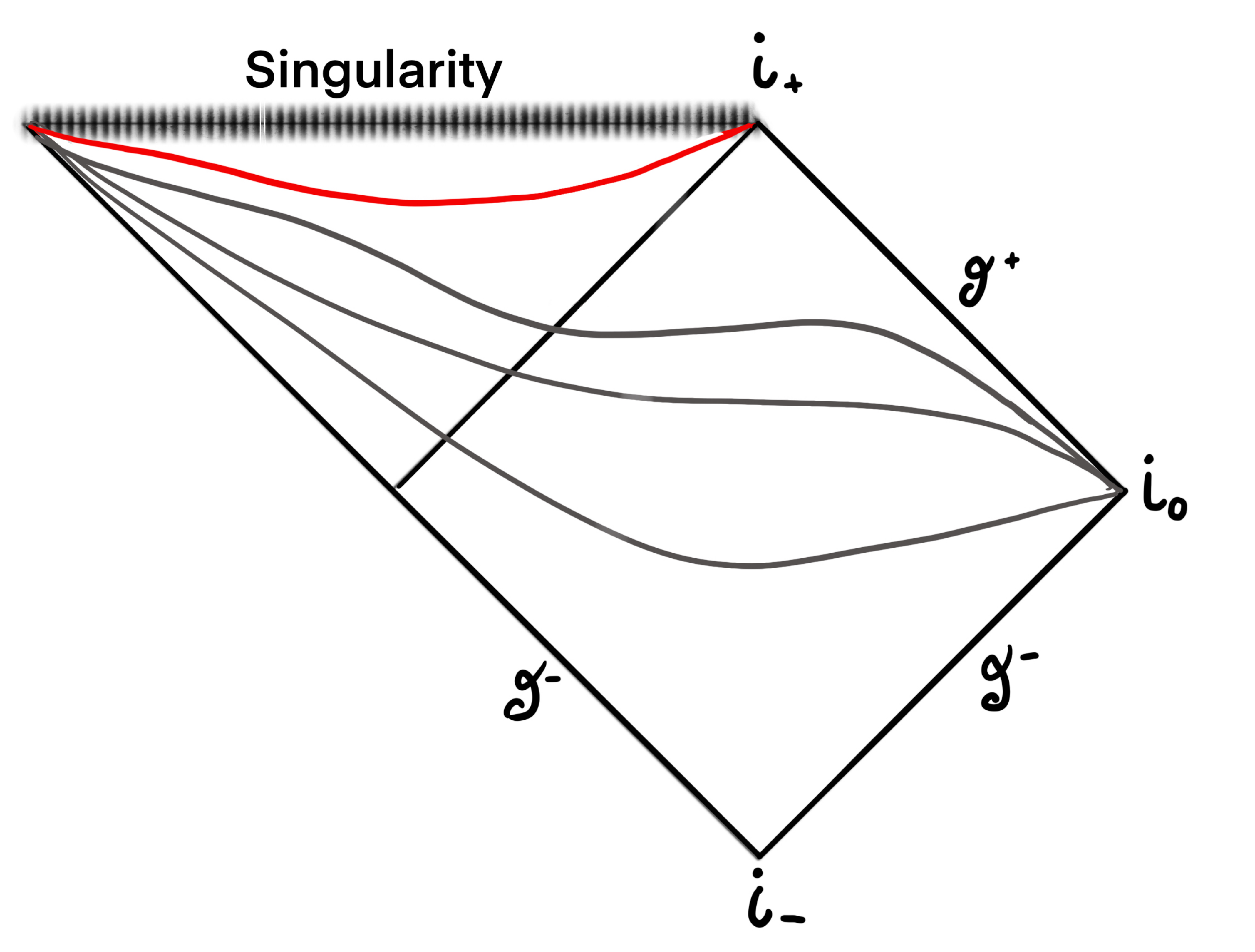}
\caption{Schematic Penrose diagram of the Lorentz violating geometry. Rays travelling at speed larger than $c$ correspond to straight lines with inclination larger than $45^{\circ}$ and can escape the Killing horizon. The gray lines represent lines of constant khronon $T$, while the red line denotes the universal horizon.}
\label{fig:penrose}
\end{figure}

However, this is the end of the similarities with the relativistic case. Due to the presence of the aether, the solution also contains a universal horizon, sitting at
\begin{align}
    U_\m \chi^\m =1-\frac{\mu}{r}=0.
\end{align}
At this radius, the integral lines of $U^\m$ wrap over themselves, becoming a compact surface of simultaneity. Moving over the $r=\mu\equiv r_{\text {\sc uh}}$ surface requires then to move at infinite speed and therefore, no signal can escape from it, thus becoming an \emph{event horizon} for any motion. Signals moving at speeds $c>1$ can penetrate the Killing horizon and eventually escape, but once they cross the universal horizon, they are trapped behind it. The schematic Penrose diagram of the geometry can be seen in figure \ref{fig:penrose}. In the following, we will work with a two-dimensional toy model by throwing away the angular coordinates and focusing on the sub-manifold defined by $t$ and $r$.

Motion within this space-time is a subtle issue. For free-falling observers moving at the speed of light, things are identical to the usual relativistic setting. They follow null trajectories of the metric \eqref{eq:metric}. However, due to its non-relativistic character, the theory also admits observers which move at different speed $c$ and therefore, in the general case, position dependent speeds $c(r)$ larger than the speed of light. 

As previously discussed, the black-hole solutions that we consider are hyper-surface orthogonal and therefore the integral curves of $U^\m$ define a global time coordinate which is shared by all observers. Thus, when writing dynamical equations, time derivatives must be identified with derivatives along $U^\mu$ while space derivatives are projected onto the orthogonal hyper-surfaces
\begin{align}
    \partial_t\equiv U^\m \partial_\m,\quad \partial_x\equiv \gamma_{\m}^{\  \n}\partial_\n,
\end{align}
where $\gamma^{\m\n}=g^{\m\n}-U^\m U^\n$ is the projector onto the hyper-surfaces orthogonal to $U^\m$. \emph{FDiff} invariance then allows for terms which contain \emph{only time derivatives or only space derivatives}, thus admitting fields with non-relativistic dispersion relations from explicit violations of LLI. One must be careful, though. Including more than two time derivatives increases the number of initial conditions required to solve the general Cauchy problem for the system and will inevitably lead to instabilities in the form of ghost degrees of freedom. Therefore, we choose to keep our action second order in time derivatives and include only higher order -- irrelevant in the IR -- operators constructed with spatial derivatives.

In the present work we will thus consider a model of a scalar field coupled to the background geometry and the aether in the following way
\begin{align}\label{eq:action_phi}
S_{\phi}=\int \di ^4x\sqrt{|g|}\,\left[\frac{1}{2}\partial_\m \phi \partial^\m \phi -\frac{2}{\Lambda^2}(\nabla_\m (\gamma^{\m\n} \nabla_\n \phi ))(\nabla_\a (\gamma^{\a\b} \nabla_\b \phi ))\right],
\end{align}
where we have included the first possible higher derivative operator which breaks Lorentz invariance, weighted by a scale $\Lambda$. The corresponding equations of motion are
\begin{align}\label{eq:eom_phi}
\nabla_\m \nabla^\m \phi+\frac{2}{\Lambda^2}(\nabla_\m \gamma^{\m\n}\nabla_\n)^2 \phi=0,
\end{align}
where the factor of $2$ is there for future convenience.

In coordinates adapted to the global time defined by the aether, where $\left|1-\frac{2\mu}{r}\right|^{\frac{1}{2}} dt_{\text{\AE}} = U_\m dx^\m$, the equations are explicitly second order in time derivatives and thus one can quantize the scalar field in the standard way, introducing modes with momentum $k^\m$ which can be decomposed as
\begin{align}\label{eq:decomposed_d}
k^\m=\omega U^\m - q S^\m,
\end{align}
and satisfy the dispersion relation, obtained in the eikonal approximation
\begin{align}\label{eq:dispersion}
\omega^2=\left(1-\frac{2\mu}{r}\right)\left[q^2+\frac{2 q^4}{\Lambda^2}\right],
\end{align}

The scale $\Lambda$ thus signals the magnitude of the momentum at which Lorentz violations become apparent in the motion of $\phi$. Note that the existence of the aether, allowing for the terms with only spatial derivatives, is crucial in order to have a modified dispersion relation.

Before going further, we should make an important remark about the quantization of these modes. In a relativistic theory, it is well known that one can construct different equivalent quantum field theories for the same action by considering different coordinate charts and thus defining the Hamiltonian by using different time coordinates. However, the theories considered here enjoy a preferred time direction, given by the integral lines along the aether field $U^\m$ as previously commented. This is important because only when the time direction is chosen to coincide with the preferred direction, the action remains second order in derivatives \cite{Blas:2009yd}.  This means that only when this time coordinate is used to define the Hamiltonian, the theory remains unitary in flat space-time. Otherwise, the action will contain up to four derivatives along any other time direction, which will lead to the presence of ghosts. Thus hereinafter we will always define the generalized momentum of the modes by using the preferred time coordinate as
\begin{align}
    \pi=U^\m \partial_\m \phi,
\end{align}
which thus always ensures the local absence of ghosts in the spectrum.

Through this work, and in order to be able to carry out an analytical analysis, we will rely on perturbation theory on $\Lambda^{-1}$, retaining the next-to-leading order contributions and in some situations, when required, also the next-to-next-to-leading order terms. Under this approximation, the dispersion relation thus becomes
\begin{align}\label{eq:cubic_disp}
    \omega=\pm\sqrt{\left|1-\frac{2\mu}{r}\right|}\left(q+\frac{q^3}{\Lambda^2}\right),
\end{align}
Being polynomial, using this dispersion relation will simplify our computation hugely. Later we will discuss the implications of retaining the full dispersion relation \eqref{eq:dispersion}.

Another important point on understanding the motion of observers in this space-time is the construction of physical free-falling trajectories, i.e the rays of massless fields coupled not only to the metric, but also to the aether and therefore endowed with a modified dispersion relation. These serve not only as the natural clocks on the geometry but also as probe observers moving under the sole influence of $g_{\m\n}$ and $U_\m$. In order to understand their motion, we consider a ray of the scalar field $\phi$ and note that its velocity vector can always be decomposed as \cite{Cropp:2013sea}
\begin{align}
V^\m=\frac{\di x^\m}{\di s}=U^\m\pm c(r) S^\m,
\end{align}
where the sign indicates whether the ray propagates outwards or inwards in the geometry and $c(r)=\frac{d\omega}{dq}$ is the group velocity of the ray. Since we are dealing with a non-relativistic theory, the precise form of $c(r)$ is frame-dependent and must be computed case by case. Note that although we have not displayed it explicitly, $c(r)$ is not only space dependent but it will also change with the energy of the ray as we will see in a moment. Specifying to $t$ and $r$, we can combine the components of $V^\m$ to get
\begin{align}
\frac{\di t}{\di r}=\frac{U^t\pm c(r)S^t}{U^r\pm c(r) S^r}=\zeta_{\pm},
\end{align}
which allows us to define the point-wise effective causal-cone coordinates of the ray as
\begin{align}\label{eq:null_coord}
\di \bar{u}=\di t-\zeta_+ \di r, \quad \di \bar{v}=\di t-\zeta_- \di r.
\end{align}
The reader can check that these reduce to the usual light-cone coordinates $u,v$ for the case of a relativistic dispersion relation. Following the integral lines of $\bar{u}$ and $\bar{v}$ is equivalent to following the trajectory of out-going and incoming rays in this geometry. At every point, the pair $(\bar u,\bar v)$ defines the causal cone of an observer communicating with signals that travel at speed $c(r)$.

At this point, it is worth noting that although the rays satisfy the modified dispersion relation \eqref{eq:dispersion} locally at every space-time point, none of the components of $k^\m$ are constant along the trajectories. Both $\omega$ and $q$ are functions of the radial coordinate $r$. Nevertheless, they can be related to conserved quantities easily. Since neither the metric nor the aether depend explicitly on the time coordinate $t$, it is easy to check that the Killing frequency $\Omega=\chi\cdot k$ is conserved. Contracting \eqref{eq:decomposed_d} with the Killing vector we thus find
\begin{align}\label{eq:solve_q}
\Omega=\omega(U\cdot \chi)-q (S\cdot \chi).
\end{align}

\begin{figure}
\begin{subfigure}{.4\textwidth}
  \includegraphics[width=.8\linewidth]{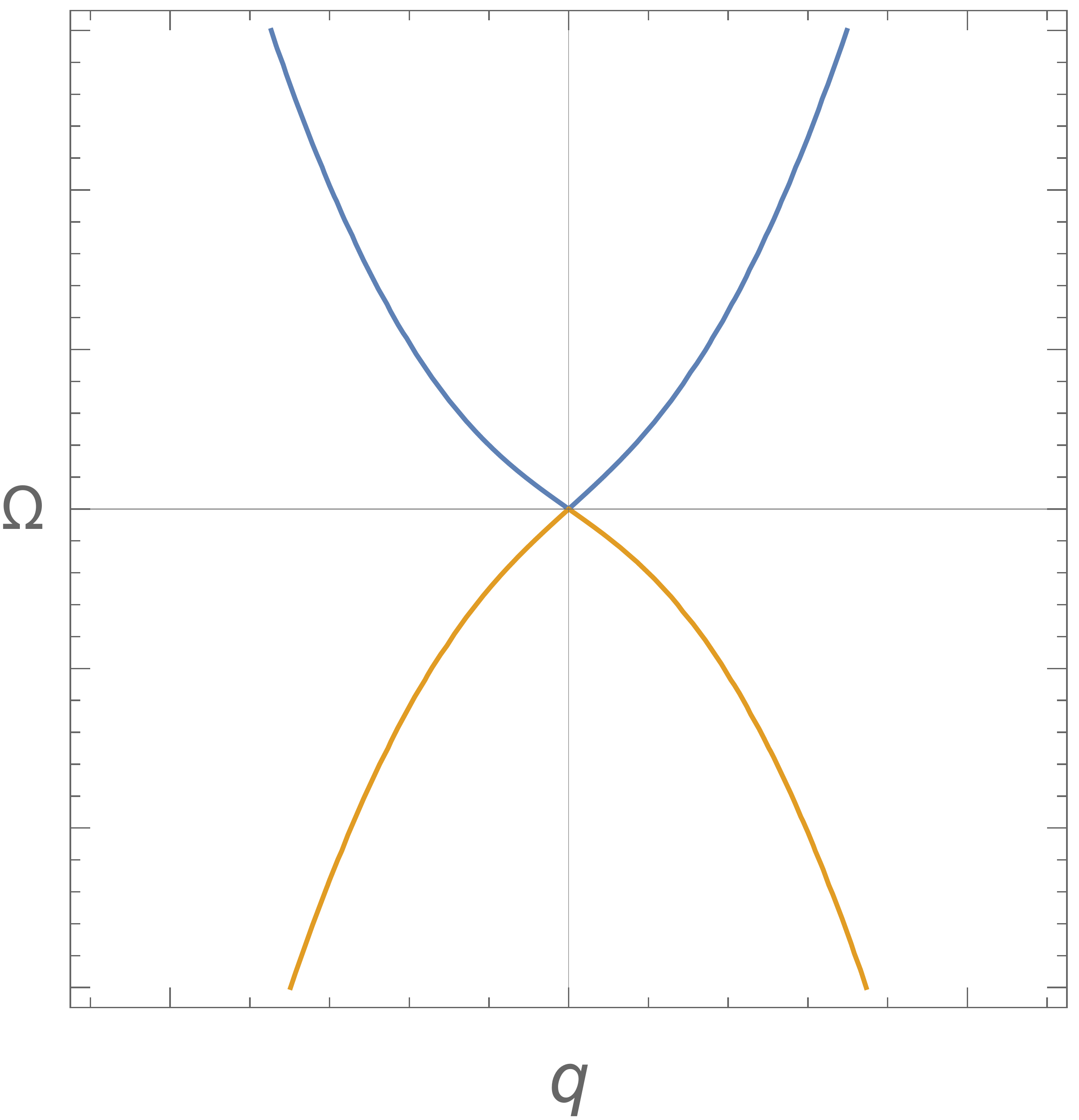}
  \caption{Branches of the dispersion relation for $r>r_{\text {\sc kh}}$.}
\end{subfigure}
\begin{subfigure}{.4\textwidth}
  \includegraphics[width=.8\linewidth]{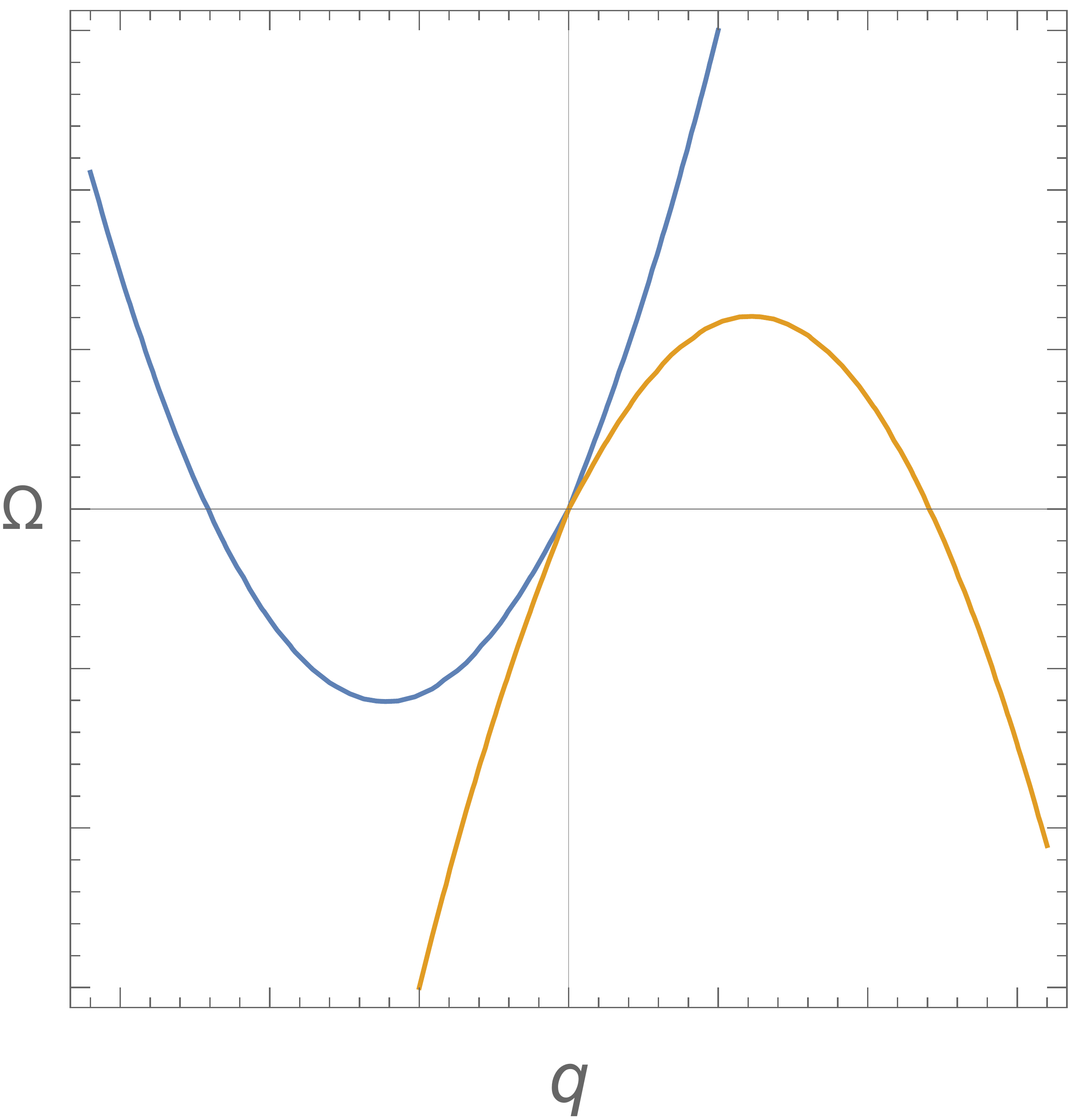}
  \caption{Branches of the dispersion relation for $r_{\text {\sc uh}}<r<r_{\text {\sc kh}}$.}
\end{subfigure}%
\caption{Schematic solution to the dispersion relation for different values of the radius. The colors distinguish the two branches in $\omega$. The blue (orange) line corresponds to the $\omega>0$ ($\omega<0$) branch. For radii larger outside the Killing horizon, there are only two solutions for positive Killing frequency $\Omega$. For radii in the region between the Universal and Killing horizons, the parabolas deform and we find four possible modes with positive $\Omega$. Two of them remain close to $|q|\sim {\cal O}(1)$ while the other two exhibit momenta comparable to the UV scale $|q|\sim\Lambda$.}
\label{fig:dispersion}
\end{figure}

\begin{figure}
  \includegraphics[width=.4\linewidth]{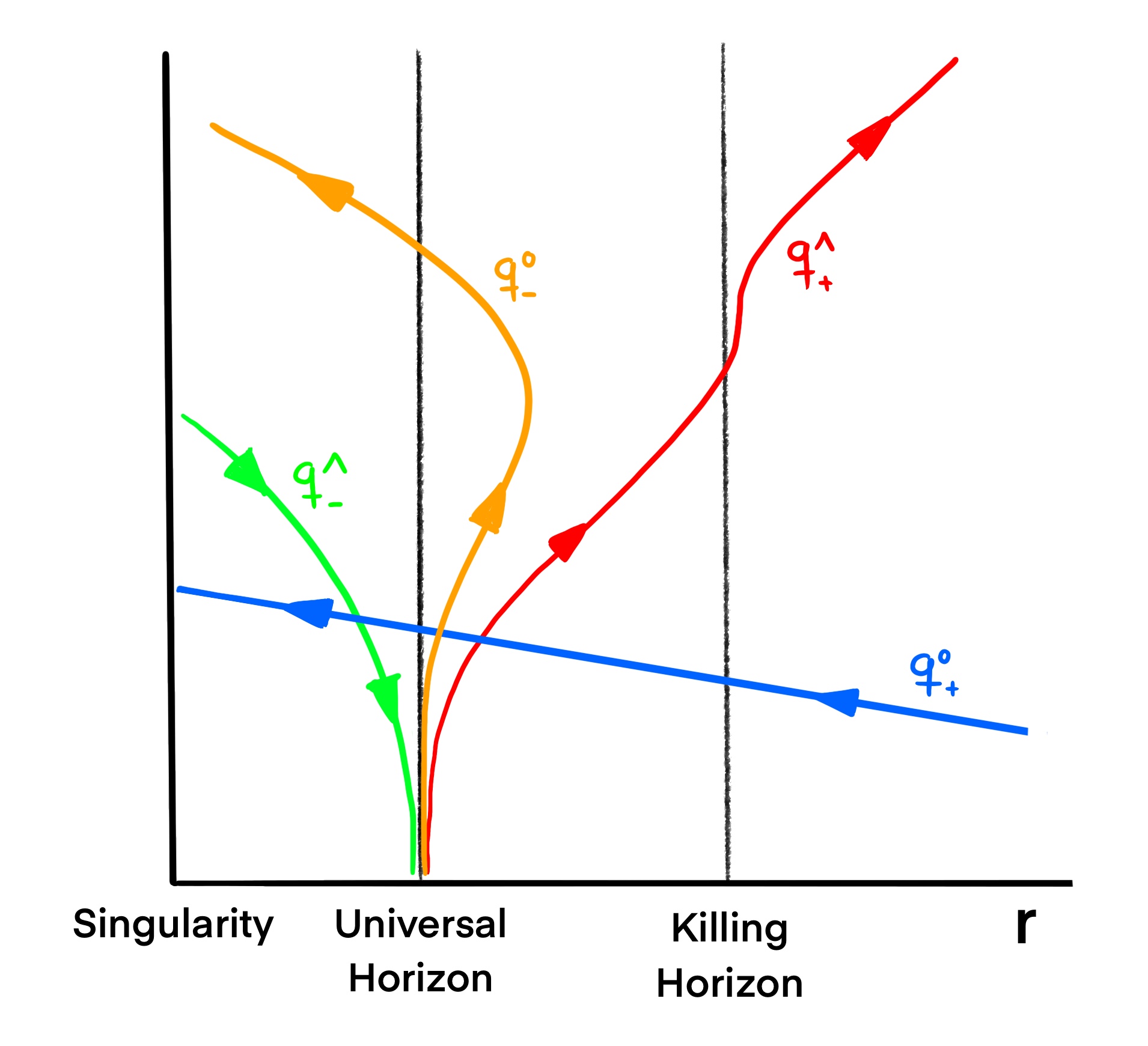}
  \caption{Artistic representation of the behavior of the modes close to the universal horizon. The proper time of every trajectory is shown by the little arrow on top of them. For a concise derivation of this behavior, check the computation in \cite{Michel:2015rsa}.}
\label{fig:modes}
\end{figure}

This allows us to study the possible solutions to the two branches of this equation -- the two branches in $\omega$ -- in terms of $r$. Their qualitative behavior is shown in figure \ref{fig:dispersion}. 
At distances larger than the Killing horizon, there exist only two solutions for positive $\Omega$, which correspond to the expected positive energy modes as given by a unitary quantum field theory. However, for radius in the range $\mu<r<2\mu$ the parabolas are deformed and cross the $\Omega=0$ axis. We thus find four positive energy solutions. In particular, for $r$ close to the universal horizon at $r_{\text {\sc uh}}=\mu$, these solutions clearly separate in two sets, whose behavior is shown in figure \ref{fig:modes}. Two of the solutions remain \emph{soft} and cross the universal horizon with finite momentum, while the other two become of order $\Lambda$. The extra two modes that now become physical correspond to the orange parabola in figure \ref{fig:dispersion} and therefore they enjoy negative aether energy and, as a consequence, negative norm in the quantum theory. However, they should not be interpreted as ghosts in the usual manner, but instead as the negative energy partners to positive modes that end up trapped behind the universal horizon. \\

The behavior of the momenta can be computed perturbatively both in $\Lambda$ and $(r-\mu)$ by choosing an appropriate ansatz and using \eqref{eq:solve_q}. For the soft modes, we take
\begin{align}
    q^{0}_{\pm}=q_0+\frac{q_1}{\Lambda}+\frac{q_2}{\Lambda^2}+\dots
\end{align}
where the subscript distinguishes the branch of the dispersion relation.

Retaining only the first sub-leading correction both for large $\Lambda$ and $r-\mu$ we find for the momenta and group velocities $c^0_{\pm}$
\begin{align}\label{eq:q0}
&q_+^{0}=\Omega\left(1+\left(2+\frac{\Omega^2}{\Lambda^2}\right)\frac{ (r-\mu)}{\mu }\right)+{\cal O}\left[\Lambda^{-3},(r-\mu)^2\right],\\
&q_-^{0}=\Omega\left(1-\frac{\Omega^2}{\Lambda^2}\frac{r-\mu}{\mu}\right)+{\cal O}\left[\Lambda^{-3},(r-\mu)^2\right],\\
&c_+^{0}=\left(1+\frac{3\Omega^2}{\Lambda^2}\right)-\left(1-\frac{9 \Omega^2}{\Lambda^2}\right)\frac{r-\mu}{\mu}+{\cal O}\left[\Lambda^{-3},(r-\mu)^2\right]\\
&c_-^{0}=\left(1+\frac{3\Omega^2}{\Lambda^2}\right)-\left(1-\frac{3 \Omega^2}{\Lambda^2}\right)\frac{r-\mu}{\mu}+{\cal O}\left[\Lambda^{-3},(r-\mu)^2\right].
\end{align}

Note that both modes are regular on the universal horizon, collapsing onto the line $\Omega=q$, which allows them to cross the horizon. We thus identify these modes as incoming modes. Following the discussion in \cite{Michel:2015rsa} it can be seen that the mode in the would-be positive branch -- the branch corresponding to positive energy outside the Killing horizon -- corresponds to a physical incoming mode while the would-be negative one represents a mode which tries to escape the gravitational well but eventually falls back.

The other modes have momentum of order $\Lambda$. Therefore we change our ansatz to
\begin{align}
    q_{\pm}^{\Lambda}=\Lambda\left( q_0 +\frac{q_1}{\Lambda}+\frac{q_2}{\Lambda^2}+\dots\right),
\end{align}
which gives, when solved perturbatively
\begin{align}\label{eq:qlambda}
    &q^{\Lambda}_{\pm}= -\Lambda\left(\pm\sqrt{\left|\frac{\mu}{r-\mu}\right|}+\frac{\Omega}{2\Lambda}+{\cal O}\left[\Lambda^{-2},(r-\mu)^1\right]\right),\\
    &c^{\Lambda}_{\pm}=\frac{3\mu}{|r-\mu|}\mp \frac{3 \Omega}{\Lambda}\sqrt{\left|\frac{\mu}{r-\mu}\right|}+{\cal O}\left[\Lambda^{-2},(r-\mu)^0\right],
\end{align}
where $c^\Lambda_{\pm}$ is the group velocity of the modes.

These modes actually diverge on the universal horizon, which prevents them from crossing it. They move infinitely close but still at a finite distance from the universal horizon. Following \cite{Michel:2015rsa} and \cite{Cropp:2013sea} we identify them with modes that linger for a long time around the Universal Horizon, thus highly blue-shifted. The would-be positive mode escapes the geometry while the would-be negative mode can be understood as incoming from the inner region and attaching to the horizon. Note that one has to be careful when analytic continuing them through $r=\mu=r_{\text {\sc uh}}$, owing to the presence of the absolute value. Finally, it is worth to point that although the mode itself is of order $\Lambda$, the solution is still under perturbative control, since it is given as an analytic expansion on $\Lambda^{-1}$ with finite and non-vanishing radius of convergence.

%%%%%%%%%%%%%%%%%%%%%%%%%%%%%%%%
%%%%%%%%%%%%%%%%%%%%%%%%%%%%%%%%
\section{Hawking's radiation from a collapsing shell: the relativistic case}
\label{sec:HRColl}
Let us for the moment forget about the existence of the aether and the modified dispersion relation and review in this section a standard approach to derive Hawking's radiation from a Schwarzschild black hole. Thus, for the moment we keep the Schwarzschild metric \eqref{eq:metric} but consider free-falling observers to be light-rays moving along null trajectories, with dispersion relation obtained in the eikonal limit
\begin{align}
\omega^2=\left(1-\frac{2\mu}{r}\right)q^2.
\end{align}

In order to derive the thermal properties of the horizon, we will follow the classic treatment by \cite{Hawking:1974sw, Parker:1975jm}. We will consider a matter configuration in form of an infinitely thin shell moving towards the center of the geometry following a time-like trajectory $r=R(\tau)$, starting from $r=R_0$, and where $\tau$ is the proper time of the shell. At early times and far enough from the shell, the space-time can be taken to be flat everywhere, and therefore one can construct the usual Minkowsky vacuum $|0\rangle_{\rm M}$ there. For later times, and at any radius $r=R(\tau)$, space-time is divided in two regions. The outer region of the shell will be described by the metric \eqref{eq:metric}, while we take the interior, without any loss of generality, to be flat
\begin{align}
\di s^2=\di \hat{t}^2 - \di r^2.
\end{align}
Eventually, the shell will cross the Schwarzschild radius $2\mu$ and will form a black hole.

We will perform an experiment of sending rays along null trajectories towards the interior of the geometry from $\mathscr{I}^-$. These rays will cross the shell, hitting its center and being re-emitted towards the exterior geometry, eventually arriving to $\mathscr{I}^+$. Incoming rays will experience a blue-shift, due to falling into the gravitational well. If the shell was static, then this blue-shift would be exactly compensated by the corresponding red-shift of the reflected ray when exiting the geometry. However, in a collapsing situation, the difference between the radius of the shell when the ray enters and exits will generate a non-negligible contribution. Our approach to compute the form of the rays (which is tantamount to the state of the modes of a field $\phi$) after the black hole has formed will be then to evaluate the shape of our probe rays after they hit $\mathscr{I}^+$. For sufficient late times, there will be a last ray which is able to enter the shell and exit it just before the horizon forms. The shape of this ray will be asymptotically close to that of the fundamental modes in the newly formed black hole geometry, from which we can obtain the flux of Hawking's radiation.

We will thus from now on follow the trajectory of a free-falling ray from $\mathscr{I}^-$ to $\mathscr{I}^+$, described in the exterior of the shell at early times by the simple function
\begin{align}
\phi(v)=\frac{e^{-i v \Omega}}{2\sqrt{\pi \Omega}},
\end{align}
where $v=t+r^*$ is the null coordinate of the metric \eqref{eq:metric}, with $r^*=r+2\mu\log\left(\frac{r}{2\mu}-1\right)$ being the tortoise coordinate, and $\Omega$ is the Killing frequency, defined as the energy of the wave at $r\rightarrow \infty$.

This ray will move towards the center of the geometry and will eventually cross the shell at $r=R(\tau)$. At that point, it will start propagating in the flat interior geometry, which will be described by a different pair of null coordinates $U,V$. In order to be able to further follow the trajectory we thus need the relation between inner and outer coordinates, $\beta(V)=v(V)$, so that the ray becomes
\begin{align}
\phi(V)=\frac{e^{-i  \Omega \beta(V)}}{2\sqrt{\pi \Omega}}.
\end{align}
Let us postpone for now the discussion of the explicit form of $\beta(V)$ and assume that we know it for the moment.

The inner null coordinates are defined by
\begin{align}
\di U=\di \hat{t} - \di r, \quad \di V=\di \hat{t} + \di r,
\end{align}
and therefore, by integrating and imposing boundary conditions on the shell we have
\begin{align}
U=\hat{t}-(r-R_0),\quad V=\hat{t}+(r-R_0).
\label{defUV}
\end{align}
This allows us to place the center of the geometry at $V=U-2 R_0$.

The ray will continue moving through the flat region until it arrives to the center of the shell at $r=0$, where it will get re-emitted. By imposing reflecting boundary conditions, the out-going wave reads
\begin{align}
\phi_R(U)=\frac{e^{-i \Omega \beta(U-2 R_0)}}{2\sqrt{\pi  \Omega}}.
\end{align}

The ray will now travel outwards and cross the shell to get out to the Schwarzschild geometry again. Defining $\alpha(u)=U(u)$, we finally have
\begin{align}
\phi_R(u)=\frac{e^{-i  \Omega \beta(\alpha(u)-2 R_0)}}{2\sqrt{\pi  \Omega}},
\label{finalmode}
\end{align}
which can be propagated to $\mathscr{I}^+$ without any additional modification. Note that $u$ corresponds to the second null coordinate in the exterior region, defined by $u=t-r^*$.

As we have seen, in order to obtain the form of the wave at $\mathscr{I}^+$ we need to compute the matching functions $\alpha(u)$ and $\beta(V)$. By differentiating with respect to the proper time of the shell, we have
\begin{align}\label{eq:dadb_GR}
\frac{\di \alpha}{\di u}=\left. \frac{\di U}{\di \tau}\frac{\di \tau}{\di u}\right|_{r=R(\tau)},\quad \left. \frac{\di \beta}{\di V}=\frac{\di v}{\di \tau}\frac{\di \tau}{\di V}\right|_{r=R(\tau)}.
\end{align}

Luckily, we are only interested in the form of the rays after the horizon forms. Thus, we can just focus on their late time behavior, when the shell gets close to the Schwarzschild radius. We thus expand its trajectory around this point,
\begin{align}
R(\tau)=2\mu + \xi (\tau_\bullet -\tau)+{\cal O}\left((\tau-\tau_\bullet)^2\right),
\label{expansion}
\end{align}
where $\tau_\bullet$ and $\xi$ are the time and (absolute value of the) speed of the shell at horizon crossing. Note that the actual speed $\frac{\partial R}{\partial \tau}=-\xi$ must be negative, indicating that the shell is collapsing. Thus $\xi>0$.

Using the metrics to obtain the differentials in \eqref{eq:dadb_GR}, the two equations can be solved in the limit $r\rightarrow 2\mu$ by keeping only the leading order contributions. We get
\begin{align}
&\alpha(u)=U(u)\sim C_1 e^{-\frac{u}{4\mu}},\\
&\beta(V)=v(V)\sim C_2+{\rm constant}\times V,
\end{align}
where the constant in front of $V$ is determined in terms of $\xi$, $\mu$, $R_0$ and $\tau_\bullet$ but its specific form is irrelevant. $C_1$ and $C_2$ are integration constants. When both solutions are combined we find that the reflected wave takes the form
\begin{align}\label{eq:result_GR}
\phi_R(u) = C_1 \exp \left(-i  \Omega C_2 e^{-\frac{u}{4\mu}}\right),
\end{align}
where we have redefined the integration constants to absorb the constant in front of $V$. As it can be seen, the wave gets an exponential red-shift, controlled by the size of the  Schwarzschild radius $2\mu$. We thus observe that while an observer in $\mathscr{I}^-$ would see vacuum --because the mode that we have sourced in the geometry is a plane wave of positive energy --, an equivalent observer sitting at $\mathscr{I}^+$ will instead observe a flux of particles coming out from the black hole. Hence, we have identified Hawking's radiation.

What is left is to determine the shape of the spectrum of the flux of particles in the radiation observed at $\mathscr{I}^{+}$. This can be done by decomposing the reflected wave onto the plane modes which serve as a basis in this region of space-time
\begin{align}
\phi_R(u)=C_1 \exp \left(-i  \Omega C_2 e^{-\frac{u}{4M}}\right)=\int_{-\infty}^\infty d\Omega' \left(\alpha_{\Omega \Omega'}\ e^{-i u \Omega'} + \beta_{\Omega \Omega'}\ e^{i u \Omega'}\right),
\end{align}
where $\alpha_{\Omega W}$ and $\beta_{\Omega W}$ are the corresponding Bogolubov coefficients.

In order to compute them, however, it is more convenient to invert the relationship and rewrite the modes defined in $\mathscr{I}^{+}$ in terms of the coordinates in $\mathscr{I}^{-}$ by using the inverse of the functions $\alpha(u)$ and $\beta(V)$ -- so that the logarithm in $\alpha^{-1}$ can be cancelled against an exponential-- and using the inner product in $\mathscr{I}^{-}$
\begin{align}
\langle \phi_1,\phi_2\rangle= i \int_{-\infty}^{\infty} dv \left(\phi_1^*\partial_v \phi_2 - (\partial_v \phi_1)^* \phi_2 \right). 
\end{align}
Things are straightforward from here and more details can be found in \cite{birrell1984quantum}. It might happen, however, in more complicated settings, that the coefficients $\alpha_{\Omega W}$ and $\beta_{\Omega W}$ contain a divergence related to the number of degrees of freedom in the radiation quanta\footnote{Indeed, if one generalizes this computation to four dimensions by including the spherical harmonics decomposition of the modes, both Bogolubov coefficients diverge, indicating that Hawking's radiation contains a steady flux of particles. Hence, the total emitted flux is infinite.}. This can be eased by noting that the divergence is shared between both coefficients and thus its ratio must be finite. In particular, we have
\begin{align}
\frac{|\alpha_{\Omega \Omega'}|^2}{|\beta_{\Omega \Omega'}|^2}=e^{8\pi M  \Omega'}.
\end{align}

This, together with the normalization condition for the modes $\int d\Omega |\alpha_{\Omega \Omega'}|^2-\int d\Omega|\beta_{\Omega \Omega'}|^2=1$, implies that the number of particles per mode in the radiation
\begin{align}
N_{\Omega'}=\int d\Omega|\beta_{\Omega \Omega'}|^2=\frac{1}{e^{8\pi M \Omega'}-1},
\label{eq:Nrel}
\end{align}
which corresponds to a Bose-Einstein distribution with temperature
\begin{align}
8\pi M  \Omega=  \frac{\Omega'}{T_{\rm H}}=\frac{2\pi  \Omega'}{\kappa}.
\end{align}

\section{Hawking's Radiation from a Universal Horizon}
\label{sec:HRUH}
We return now to the situation of a black hole in Einstein-Aether Gravity, where we follow the motion of a field with an anisotropic dispersion relation \eqref{eq:dispersion} in the aether frame. Our aim in the following will be to reproduce the previous computation of the horizon temperature in this case.

We will consider, as in the previous section, an infinitely thin shell following a time-like trajectory $r=R(\tau)$ towards the center of the geometry\footnote{Notice that this implies that, unlike the scalar field $\phi$, the shell is made of ordinary matter with a Lorentz invariant dispersion relation, thus decoupled from the aether.}. However, in this case, and unlike in the relativistic case, we will not stop tracking the shell after it reaches the Killing horizon. The reason is that now observers and rays travelling at speeds larger than the speed of light are allowed. While in a relativistic setting any ray that crosses the Killing horizon is forever trapped within its interior, that is not the case anymore. Here we have rays that move at arbitrary speed, so the process of crossing the Killing horizon and escaping from its interior is possible. Instead, it is the universal horizon which serves as an event horizon trapping any ray in its interior. Therefore this will be the ultimate region down to which we can follow the evolution of the shell. After the shell crosses the universal horizon at $\tau=\tau_\bullet$, we will consider that the black hole is formed. Note also that the aether field is completely regular at the universal horizon, when $r=\mu$, laying along the timelike direction of space-time. Thus, we can continuously glue a flat interior to the shell without any discontinuity. A deeper analysis of this junction can be found in \cite{Michel:2015rsa}\footnote{Let us stress however, that due to the absence of an equivalent of Birkhoff's theorem for Einstein-Aether gravity, the existence of this solution is a reasonable assumption, but it has yet to be formally proven by dynamical evolution of a spherical collapsing space-time. The formation of the universal horizon, though, has been shown in some particular scenarios ~\cite{Saravani:2013kva,Franchini:2021bpt}.}.

Another important difference to discuss here are the features of the asymptotic null infinity regions for incoming and out-coming rays, before and after crossing the shell. Indeed, high energy rays can have superluminal group velocities and in principle start in the past, and reach in the future, not only the low energy null asymptotics $\mathscr{I}^{\mp}$ respectively, but they can also start and reach spatial infinity $i^0$. We shall hence construct our modes using the previously introduced effective causal coordinates from \eqref{eq:null_coord} which include the possibility of superluminal motion and hence of incoming rays from spatial infinity. For what regards the outgoing ones we shall see {\em a posteriori} that the spectrum of rays that arrive at large radii after crossing the shell is exponentially red-shifted at late times, thus peaked at low energies as usual. The propagation of these rays will then be close to the general relativistic one, and therefore $\mathscr{I}^{+}$ will be approximately equal for all rays.

We start by describing the incoming ray at $\mathscr{I}^-$. In this region, the energy of the wave corresponds to the Killing frequency. Using the effective causal coordinates from \eqref{eq:null_coord}, the incoming wave reads
\begin{align}
    \phi(v)=\frac{e^{-i \bar v \Omega}}{2\sqrt{\pi \Omega}}.
\end{align}
Again, in the asymptotic region $r\rightarrow \infty$ space-time is flat and we can construct the usual Minkowsky vacuum there $|0\rangle_{\rm M}$.

As in the relativistic case, this wave travels inwards towards the shell following the effective causal coordinate, i.e. a free-falling trajectory. Once it hits the surface of the shell, however, it will start propagating along the corresponding coordinate in the inner flat space. Afterwards, it will bounce back in the center of the geometry and, after crossing another portion of flat space, it will escape the shell and travel up to $\mathscr{I}^+$. In an identical way to the relativistic case, the reflected wave, once it arrives to $\mathscr{I}^+$, will read
\begin{align}
    \phi_R(\bar u)=\frac{e^{-i \Omega \beta(\alpha(\bar u)-2 \bar{c} R_0)}}{2 \sqrt{\pi \Omega}}.
\end{align}
The coefficient $\bar{c}$ denotes the fact that the waves propagate at a speed $\bar{c}$ different from the speed of light within the inner flat geometry. The corresponding causal coordinates are thus now
\begin{align}
    \bar U=\hat t - \bar{c}(r-R_0), \quad \bar V=\hat t +\bar{c}(r-R_0),
\end{align}
and the center of coordinates corresponds to $\bar V=\bar U-2\bar{c}R_0$. The concrete value of $\bar{c}$ could be determined by the speed $\bar{c}=\left.c(q)\right|_{r=R(\tau)}$ of the wave at shell-crossing but, being constant, its particular value is irrelevant for our computation here.

We are thus left with the task of computing the matching functions $\alpha(\bar u)$ and $\beta(\bar V)$ between the inner and outer coordinates. As in the relativistic case, we can differentiate with respect to the proper time of the shell $\tau$, getting
\begin{align}
    \frac{d\alpha}{d\bar u}=\left.\frac{d\bar U}{d\tau}\frac{d\tau}{d\bar u}\right|_{r=R(\tau)},\quad \frac{d\beta}{d\bar V}=\left.\frac{d\bar v}{d\tau}\frac{d\tau}{d\bar V}\right|_{r=R(\tau)}.
\end{align}

Let us first focus on $\alpha(u)$. From the metrics and the explicit form of the inner null coordinates we get
\begin{align}
    &\frac{d\bar U}{d\tau}=\sqrt{1+\dot{R}^2}-\bar{c}\dot{R},\\
    &\frac{d\bar u}{d\tau}=\sqrt{\frac{1}{F(R)}+\frac{\dot{R}^2}{F(R)^2}}\ - \zeta_+ \dot{R},
\end{align}
so that
\begin{align}\label{eq:alphap}
    \frac{d\alpha}{d\bar u}=\frac{\sqrt{1+\dot{R}^2}-\bar{c}\dot{R}}{\sqrt{\frac{1}{F(R)}+\frac{\dot{R}^2}{F(R)^2}}\ - \zeta_+ \dot{R}},
\end{align}
where $\dot{R}=\frac{dR}{d\tau}$.

Let us note a striking difference with respect to the relativistic case. Since the ray is accelerating and enters the shell with different speeds at different time snapshots, we find an explicit dependence in the group velocity, contained within $\zeta_+$, which changes with $r$. Since we are interested in the late time epoch, when the shell is asymptotically close to the universal horizon, we will only need to evaluate the behavior of $\zeta_+$ in this region. In order to do that, we need the relation between the momentum $q(r)$ and the Killing frequency $\Omega$ in \eqref{eq:solve_q} for the particular mode that we are tracking. Since $\alpha(\bar u)$ is the matching function for the coordinates following the trajectory of the out-going ray, after being reflected at the center of coordinates, we need to insert here the group velocity $c(r)=\frac{d\omega}{dq}$ for out-going rays, that we can evaluate in the overlap region when the reflected ray approaches the shell at $r\sim\mu+ 0^{-}$. Thus, from \eqref{eq:qlambda} we get
\bea
    &\zeta_+=\frac{3}{2}\frac{\mu}{|r-\mu|}+\frac{3\Omega}{4\Lambda}\sqrt{\left|\frac{\mu}{r-\mu}\right|}+{\cal O}\left[\Lambda^{-2},(r-\mu)^0\right].
\eea

Going back to \eqref{eq:alphap} and expanding the trajectory of the shell close to the universal horizon at leading order in $(r-\mu)$, which corresponds to leading order in $(\tau-\tau_\bullet)$, we get
\begin{align}
    R(\tau)=\mu + \xi (\tau_\bullet -\tau)+{\cal O}\left((\tau-\tau_\bullet)^2\right),
\end{align}
so that we finally have
 \begin{align}
    &\frac{d\alpha}{d\bar u}=\frac{d\bar U}{d\bar u}=\frac{2}{3\mu \xi}\left(\sqrt{1+\xi^2}+\bar{c}\xi\right)(R(\tau)-\mu)+{\cal O}\left[\Lambda^{-3},(r-\mu)^{3/2}\right],
\end{align}
where we have evaluated $\dot{R}=-\xi$ and we remind that $\bar{c}$ is a constant.

The last step is to find a relation between $R(\tau)$ and $\bar U$ in order to have a differential equation that we can solve. This can be done easily by using the shell trajectory and the metric
\begin{align}
    \frac{dR}{d\bar U}=\frac{dR}{d\tau}\frac{d\tau}{d\bar{U}}\sim\frac{\dot{R}}{\sqrt{1+\dot{R}^2}-\bar{c}\dot{R}},\longrightarrow R\sim -\frac{\xi U}{\sqrt{1+\xi^2}+\bar{c}\xi}+C_0,
\end{align}
with $C_0$ an integration constant. With this we finally get
\begin{align}
    \frac{d\alpha}{d\bar u}=\frac{d\bar U}{d\bar u}=\frac{2}{3\mu \xi}\left(\sqrt{1+\xi^2}+\bar{c}\xi\right)\left(-\frac{\xi U}{\sqrt{1+\xi^2}+\bar{c}\xi}+C_0-\mu\right)+{\cal O}\left[\Lambda^{-3},(r-\mu)^{3/2}\right].
\end{align}

This can now be easily integrated to yield
\begin{align}\label{eq:alpha}
    \alpha(\bar u)=\bar U(\bar u)=C_1 e^{-\frac{2\bar u}{3\mu}}+C_0+{\cal O}\left[\Lambda^{-3},(r-\mu)^2\right],
\end{align}
where $C_1$ is another integration constant and we have redefined $C_0$ to absorb constant terms into it.

We turn now to computing $\beta(\bar V)$. Using again the metrics, we find that the differential equation governing the behavior of $\beta(V)$ can be written as
\begin{align}
    \frac{d\beta}{d\bar V}=\frac{\sqrt{\frac{1}{F(R)}+\frac{\dot{R}^2}{F(R)^2}}-\zeta_- \dot{R}}{\sqrt{\dot{R}^2+1}+\bar{c}\dot{R}}.
    \end{align}
In this case, we must take into account that $\beta(\bar V)$ is the matching function for the coordinates of the ray incoming into the shell. Therefore, we must compute $\zeta_-$ by evaluating the mode $q_+^0$ in \eqref{eq:q0}, which gives
\begin{align}
\zeta_-=1 + \frac{3 \Omega^2}{\Lambda^2}+ \left(1+ \frac{9  \Omega^2 }{\Lambda^2} \right) \dfrac{(r-\mu)}{\mu} +{\cal O}\left[\Lambda^{-3},(r-\mu)^2\right].
\end{align}

We plug this together with the expansion of the trajectory of the shell close to the universal horizon and we find
\begin{align}
    \frac{d\beta}{d\bar V}=\frac{d\bar v}{d\bar V}=\frac{1}{-\bar{c}\xi+\sqrt{1+\xi^2}}\left[\sqrt{\xi^2-1}+\xi +\frac{3\xi\Omega^2}{\Lambda^2}\right]+{\cal O}\left[\Lambda^{-3},(r-\mu)\right],
\end{align}
which trivially integrates to
\begin{align}
    \beta(\bar V)=\bar v(\bar V)=\frac{\bar V}{-\bar{c}\xi+\sqrt{1+\xi^2}}\left[\sqrt{\xi^2-1}+\xi +\frac{3\xi\Omega^2}{\Lambda^2}\right]+C_2+{\cal O}\left[\Lambda^{-3},(r-\mu)\right],
\end{align}
where $C_2$ is an integration constant.

Collecting both solutions together we thus find that the reflected wave takes the final form
\begin{align}
    \phi_R(\bar u)=C_2 \exp\left(-i C_1 \Omega e^{-\frac{2\bar u}{3\mu}}\right),
\end{align}
where we have redefined the integration constants, absorbing factors of $\xi$, $\bar{c}$, $\Omega$ and $\Lambda$ into them.

As in the relativistic case, we find that the reflected wave gets exponentially red-shifted by a quantity controlled by the position of the universal horizon $r_{\text{\sc uh}}=\mu$ with an inverse power law. However, the numerical factor does not coincide with the one derived in General Relativity, although both are ${\cal O}(1)$. In the next section, we will discuss this numerical difference, tracing it back to the form of the high energy correction dominating the dispersion relation for large $q$.

The final step is to project this reflected wave onto the eigenstates of positive and negative energy in $\mathscr{I}^+$. As shown in figure \ref{fig:dispersion}, for large radii we only have two possible solutions for the dispersion relation, which take a form analogous to the relativistic case
\begin{align}
    \phi(\bar{u})^{\pm}_{\mathscr{I}^{+}}=\frac{e^{\pm i \Omega' \bar{u}}}{2\sqrt{\pi \Omega'}}.
\end{align}

Owing to this fact, we can straightforwardly borrow the result from the relativistic case and read the universal horizon temperature from our result to be
\begin{align}\label{eq:HT}
    T_{\text{\sc uh}}=\frac{1}{3\pi \mu}=\frac{\bar{\kappa}}{2\pi}.
\end{align}

Note however that in this case $\bar \kappa=\frac{2}{3\mu}$ does not correspond to the surface gravity at the universal horizon, unlike what happens in a relativistic setting, where the surface gravity of the Killing horizon controls the temperature. This can be easily confirmed by computing $\kappa_{\text{\sc uh}}$ following the prescription in \cite{Cropp:2013sea,Cropp:2013zxi} (appropriately modified for our different signature here)
\begin{align}
\kappa_{\text{\sc uh}}=-\left. \frac{1}{2}U^\a \nabla_\a \left(U^\m \chi_\m\right)\right|_{r_{\text {\sc uh}}}=\frac{1}{2\mu},
\label{eq:kappaUH}
\end{align}

With this knowledge we can write our result for the $\bar \kappa$ governing Hawking's radiation as
\begin{align}
    \bar\kappa=\frac{4\kappa_{\text{\sc uh}}}{3}.
\end{align}

\section{Overlapping of WKB modes nearby the Universal Horizon}
\label{sec:WKB}
We offer here an alternative derivation of the temperature of the Hawking radiation just obtained in the previous section. We will follow the steps of \cite{Michel:2015rsa}, aiming to describe the WKB modes of \emph{out-going} radiation close to the collapsing shell. These modes correspond to the natural plane waves in this region of the space-time. Afterwards, we will compute their inner product with a plane wave annihilating the Minkowsky vacuum inside the shell. By taking the late-time limit, as before, this will give an asymptotically good approximation to the overlapping of modes once the universal horizon has formed, allowing us to derive the thermal properties of the radiation, if any.

The main difference between \cite{Michel:2015rsa} and our work here will be the choice of frame to construct WKB modes of the radiation close to the universal horizon. Since WKB modes are not diffeomorphism invariant, and this theory is not either, there is no reason as to why modes constructed in different frames must lead to the same Bogoliubov coefficients once projected onto flat waves. This of course opens a conundrum, the choice of frame for describing the modes corresponds to different observers moving close to the universal horizon with different kinematics. Since, as we have seen, this property is crucial for determining the thermal flux of the universal horizon, one can {\em a priori} expect the result to depend strongly on the chart of coordinates used when approaching the universal horizon. In \cite{Michel:2015rsa} modes described in the aether frame were considered, thus depending on the preferred coordinates in this frame, which can be extended towards the inner region. Here instead, we choose to follow the generalization of free-falling observers to the case of a modified dispersion relation \eqref{eq:dispersion}. Thus, we need to describe our WKB modes in the proper frame of these observers.

We start by performing a change of coordinates from $dt$ to $d\bar u=dt-\zeta_+ dr$ in order to adapt the coordinate system to the propagation of modes outgoing from the universal horizon. Although the expression for $\bar u$ might be complicated enough that it cannot be obtained analytically, we only need to perform the change on the metric and the aether vector, which depend only on $r$ and the differentials.

We now go back to the decomposition of the momentum vector \eqref{eq:decomposed_d} and contract it with the generator of the spatial radial coordinate $\partial_r$ in order to get the spatial momentum $P$ in this frame. We thus have
\begin{align}
    P(r)=q+\frac{\mu}{r}(\omega-q).
\end{align}

As previously discussed, we are interested in the shape of outgoing radiation. Note however that in this case we also need to take into account the branch of the solutions with negative energy, since we need to consider the whole vector space that the four modes span.

We now construct WKB modes in this region from the expression
\begin{align}
    \Psi_{\rm WKB}= \frac{\exp\left[-i \left(\Omega' u -\int dr P(r)   \right)\right]}{4\pi \sqrt{|\Omega'/(c(r) \partial_{\Omega'} P)|}},
    \label{WKBdef}
\end{align}
where $c(r)$ and $P(r)$ must be plugged case by case for every mode. Here we use $\Omega'$ to denote the killing energy of these modes since this will be the base onto which we are projecting.

Going back to figure \ref{fig:dispersion} we remind that there exist four possible modes propagating close to the universal horizon, which correspond to the soft modes, say $\eta_{\pm}$, and the hard modes, which we shall call $\Psi_{\pm}$. The  $\pm$ denotes whether they correspond to the blue (orange) curve, identifying positive (negative) energy in the aether frame. Thus, in the region outside the universal horizon, any perturbation $\phi(u,r)$ with Killing frequency $\Omega'$ can be decomposed in a basis of these modes
\begin{align}
    \phi(u,r)=\int_{-\infty}^{\infty} d\Omega \left(\alpha_{\Omega \Omega'} \Psi_+ +\beta_{\Omega \Omega'} \Psi_- + \delta_{\Omega \Omega'} \eta_+ +\gamma_{\Omega \Omega'} \eta_-  \right),
\end{align}
where the $\pm$ denotes if they correspond to the branch with positive or negative energy.

As before, for those modes annihilating the Minkowsky vacuum, we take plane waves with Killing frequency $\Omega'$
\begin{align}
    \phi_{M}(u)=\frac{e^{-i U \Omega}}{2\sqrt{\pi \Omega}},
\end{align}
which we interpret here as the modes in the flat interior of the shell. These are the modes that we want to compare with those on the other side of the shell. 

In order to evaluate the late-time behavior of the Bogoluvob coefficients, we can take into account that the shell is a closed surface, which implies that only modes with a certain frequency, inversely proportional to the radius $R(s)$ of the shell, are possible. Thus, asymptotically late times can be identified onto the limit $\Omega \rightarrow \infty$. Finally, since we want to evaluate them in the exterior of the shell, we need to perform a change of variables involving the matching function $\alpha(u)=U(u)$ in \eqref{eq:alpha}.

The causal inner product of $\phi_{M}(u)$ with $\eta_{\pm}$ can be easily shown to vanish. Close to the universal horizon, the modes are regular and a standard dispersion relation $|q|=\Omega'$ is recovered. Therefore, only soft positive modes overlap between themselves, leading to no radiation. We are thus left with the task of evaluating the overlap with the hard modes $\Psi_{\pm}$. Their momentum $P_{\pm}(r)$ can be easily computed perturbatively in $\Lambda$ to get 
\begin{align}
    P_{\pm}(r)= -\left(\Lambda \left(\frac{\mu}{r-\mu}\right)^{\frac{3}{2}}\pm\frac{3}{2}\frac{\Omega' \mu}{(r-\mu)} \right)+{\cal O}\left[(r-\mu)^{-\frac{1}{2}} \right],
\end{align}
where the sub-leading term distinguishes the two $\pm$ roots and we need 
to keep it in order for the derivative $\partial_{\Omega'} P_{\pm}$ not to vanish. Note that the pre-factor that controls $\bar \kappa$ appears in front of the term $(r-\mu)^{-1}$ which becomes a logarithm upon integration to construct the WKB mode. Indeed, if we compare this with $q^{\Lambda}$ in \eqref{eq:qlambda} we see that no term with a $(r-\mu)^{-1}$ arises for the momentum in the aether frame. This hints that the logarithm must take an important role in providing the thermal spectrum here.

We thus have two WKB modes
\begin{align}
    \Psi_\pm=\frac{3\mu}{4\pi(r-\mu)\sqrt{2\Omega'}}\times \exp \left[-i\left(\Omega' u -\frac{2 \Lambda \mu^{\frac{3}{2}}}{\sqrt{r-\mu}}\pm\frac{3\Omega' \mu}{2}\log \left|\frac{r-\mu}{\mu}\right|\right)\right]
\end{align}
and we focus on computing their inner product with the flat modes of positive energy $\phi_M$ in the proximity of the universal horizon
\begin{align}
    \langle \phi_M, \Psi_{\pm}\rangle= i \int S_\m dx^\m \left( \phi^*_M \Pi_{\pm}+\Psi_{\pm} \pi^*_M \right),
\end{align}
where $\Pi_{\pm}=U^\m \partial_\m \Psi_{\pm}$ and $\pi_M=U^\m\partial_\m \phi_M$ are the momenta associated to the modes, derived from the action \eqref{eq:action_phi}.

At the leading order the integration measure is just
\begin{align}
    S_\m dx^\m= \frac{3\mu}{2(r-\mu)}dr+{\cal O}\left[(r-\mu)^0\right]
\end{align}
so that we get
\begin{align}
\nonumber \langle \phi_M, \Psi_\pm\rangle&= \frac{9 i \Lambda }{32\pi^2 \sqrt{2 \Omega \Omega'}}e^{-i\left(\Omega' u - C_1 \Omega e^{-\frac{2 u}{3 \mu}}\right)} \int_\mu^\infty dr\ \frac{\mu^{7/2}}{(r-\mu)^{7/2}}\exp \left[-i\left(-\frac{2 \Lambda \mu^{\frac{3}{2}}}{\sqrt{r-\mu}}\pm\frac{3\Omega' \mu}{2}\log\left|\frac{ r-\mu}{\mu}\right|\right)\right]\\
     &=A \int_\mu^\infty  \frac{dr}{(r-\mu)^{7/2}}\exp \left[-i\left(-\frac{2 \Lambda \mu^{\frac{3}{2}}}{\sqrt{r-\mu}}\pm\frac{3\Omega' \mu}{2}\log \left|r-\mu\right|\right)\right],
\end{align} 
where we have enclosed the different pre-factors into the constant $A$.

The computation of this integral is not trivial, due to the branch cut of both the square root and the logarithm. One of the integration limits lays on the branch point and therefore the exponent within the integrand is not a holomorphic function along the integration regime, which forbids us from applying a saddle point approximation. One could in principle try to regularize this by shifting the integration regime to $\{\mu+\epsilon,\infty\}$ and then taking the limit of $\epsilon$ to vanish. However, this fails, since the integral does not converge for real values of $\Lambda$ and $\mu$. Instead, an approximate formula which is enough for our purposes here can be obtained by deforming the integration contour.

\begin{figure}
  \includegraphics[scale=.3]{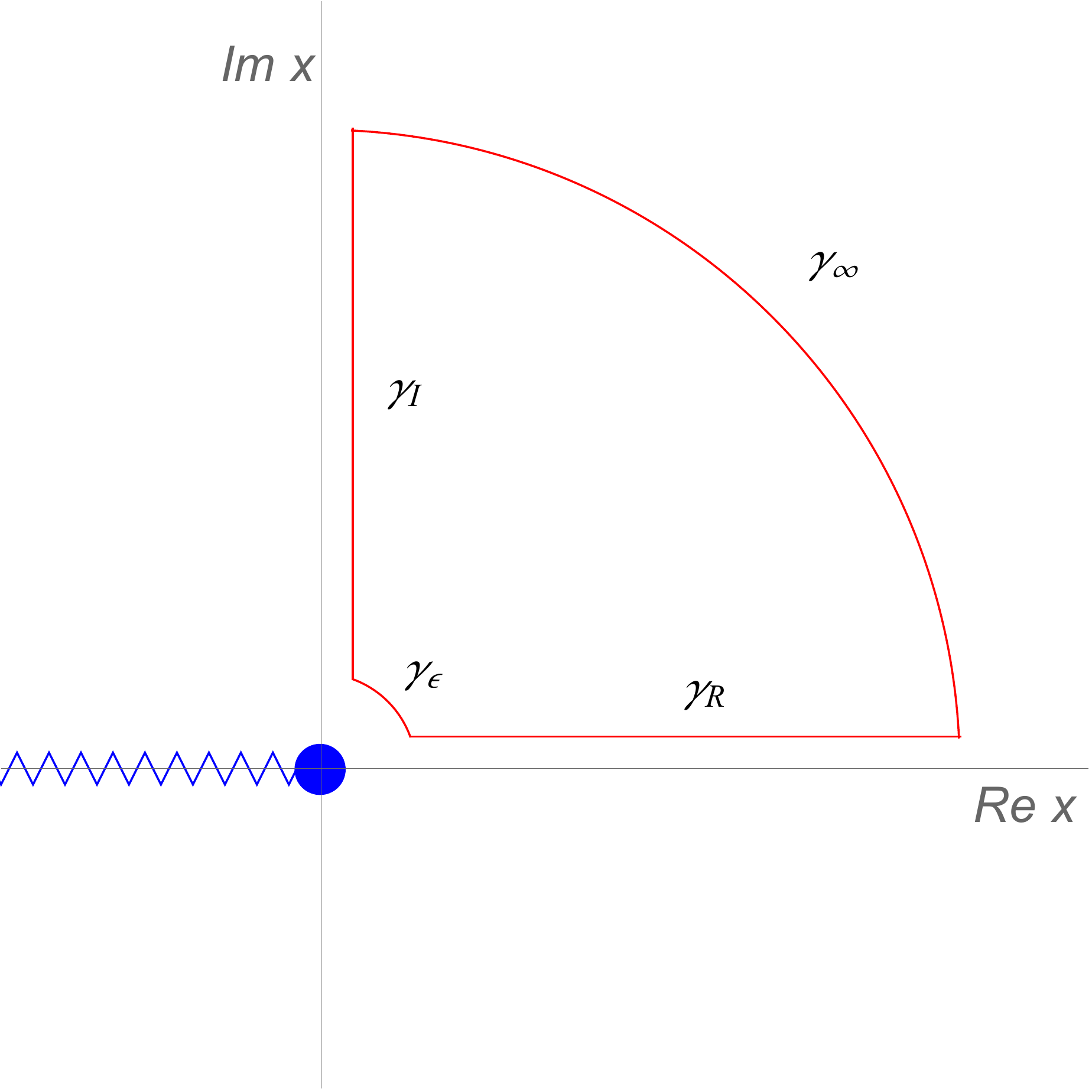}
\caption{Integration contour $\gamma$ in the complex plane. The outer circumference has radius $|x|\rightarrow \infty$, while the inner one is described by the curve $x=\epsilon\ e^{i\theta}$, with $0<\theta<\pi/2$. The branch cut is shown in blue.}
\label{fig:contour}
\end{figure}

First, we shift the integration variable to $x=r-\mu$ so that we attach the branch point to $x=0$ and we choose to define the integrand through its principal value, with the branch cut running along the negative real axis in the complex plane. We then consider the integral 
\begin{align}
     \Delta_\pm=A \oint_\gamma  \frac{dx}{x^{7/2}}\exp \left[-i\left(-\frac{2 \Lambda \mu^{\frac{3}{2}}}{\sqrt{x}}\pm\frac{3\Omega' \mu}{2}\log |x|\right)\right],
\end{align}
where the contour $\gamma=\gamma_R+\gamma_\infty +\gamma_I+\gamma_\epsilon $ is depicted in figure \ref{fig:contour}.

Using Cauchy's integral theorem, we find that $\Delta_{\pm}=0$. On the other hand, the integral over $\gamma_{\infty}$ can be shown easily to vanish, while that over $\gamma_{\epsilon}$ does too if we close the contour through the upper half of the complex plane, as shown in figure \ref{fig:contour}. We are thus left with
\begin{align}
     \Delta_\pm=-A \int_{\epsilon}^\infty \frac{dx}{x^{7/2}}\exp \left[-i\left(-\frac{2 \Lambda \mu^{\frac{3}{2}}}{\sqrt{x}}\pm\frac{3\Omega' \mu}{2}\log |x|\right)\right]+A \int_{i\infty}^{i\epsilon} \frac{dx}{x^{7/2}}\exp \left[-i\left(-\frac{2 \Lambda \mu^{\frac{3}{2}}}{\sqrt{x}}\pm\frac{3\Omega' \mu}{2}\log |x|\right)\right]=0,
\end{align}
which implies
\begin{align}
    \langle \phi_M, \Psi_\pm\rangle&=A \int_{i\infty}^{i\epsilon} \frac{dx}{x^{7/2}}\exp \left[-i\left(-\frac{2 \Lambda \mu^{\frac{3}{2}}}{\sqrt{x}}\pm\frac{3\Omega' \mu}{2}\log |x|\right)\right].
\end{align}

We now perform another change of variables $x=iy$, so that we get
\begin{align}
  \nonumber   \langle \phi_M, \Psi_\pm\rangle&= -i A  \int_{\epsilon}^{\infty} \frac{dy}{(i y)^{7/2}}\exp \left[-i\left(\frac{\sqrt{2}(1-i) \Lambda \mu^{\frac{3}{2}}}{\sqrt{y}}\pm\frac{3\Omega' \mu}{2}\log (y)\right)\right].
\end{align}

Note that now this integral is more convergent that the original one when $\epsilon\rightarrow 0$, due to the real contribution to the exponential from the square root. Indeed, it now converges for positive $\epsilon$ as long as both $\Lambda$ and $\mu$ are positive. Using the analytic extension of the Gamma function we get
\begin{align}
    |\langle \phi_M, \Psi_\pm\rangle|^2&\propto |A|^2  \left|\Gamma[5-3i \mu \Omega']\right|^2 e^{\pm \frac{3\pi \Omega' \mu}{2}},
\end{align}
where the proportionality factor is the same for both $\Psi_{\pm}$.

Therefore we have that the ratio between the Bogoluvob coefficients becomes
\begin{align}
     \frac{|\alpha_{\Omega W}|^2}{|\beta_{\Omega W}|^2}=   \frac{|\langle \phi_M, \Psi_+\rangle|^2}{|\langle \phi_M, \Psi_-\rangle|^2}=e^{3\pi\Omega'\mu}+\dots,
\end{align}
where the dots stand for sub-leading terms in $r-\mu$. 

So finally, in strict analogy with the derivation of Eq.~\eqref{eq:Nrel}, we can use the above result and the normalization condition for the Bogoliubov coefficients to deduce that the spectrum observed at $\mathscr{I}^+$ will be Planckian with temperature $T_{\text{\sc uh}}={1}/{(3\pi \mu)}={\bar{\kappa}}/{(2\pi)}$. 
Noticeably, this result agrees with expression \eqref{eq:HT} obtained from the alternative derivation based on mode-matching for a collapsing shell.

An obvious question arises here. Why are we finding a thermal spectrum coming from the universal horizon while the authors of \cite{Michel:2015rsa} did not find such an effect? In our language here the reason is clear. The temperature of the radiation is controlled by the $(r-\mu)^{-1}$ term in $P_{\pm}(r)$, which integrates to a logarithm in the exponent of the WKB mode. This term only appears once we change frame and construct these WKB modes in the frame co-moving with free-falling observers or physical rays of the field $\phi$. This choice was motivated by the fact that the quantum modes of the field correspond indeed to avatars of physical rays and therefore this seems to be the natural frame to follow their dynamics on this space-time. However, this choice is of course not the unique one. Instead, one could decide to stick to the preferred frame given by the aether time and its orthogonal hyper-surfaces, which corresponds to the choice made by \cite{Michel:2015rsa}. In that case\footnote{Note however that our setting here, although physically equivalent, is technically not identical to the one considered in \cite{Michel:2015rsa}. In there, they indeed have a term $(r-\mu)^{-1}$ in the momentum of the preferred frame, but their computation is slightly different and eventually leads to the same physical conclusions.}, however, note that the momenta that we need to insert in order to construct the WKB modes are the $q^\Lambda_{\pm}$, which do not have any $(r-\mu)^{-1}$ term. Instead, around the universal horizon, these WKB modes would collapse onto flat modes, since 
\begin{align}
    \int dr \, q^\Lambda_\pm =-\frac{\Omega' \mu}{2}+{\cal O}(r-\mu)^{1/2}, 
\end{align}
from which is trivial to check that no radiation is obtained.

In summary, we can say that the difference between our framework and that of \cite{Michel:2015rsa} is tantamount to a different choice of the vacuum state at the universal horizon which, due to the non-relativistic causality structure of the theory, is not fully fixed by the choice of vacuum state just on $\mathscr{I}^-$. In our case, the state is supposed to be the vacuum for physical rays of the field, while in the case of \cite{Michel:2015rsa} it was chosen to be that for the aether frame. Of course, we expect that if one would be able to set up a well posed Cauchy problem for a collapsing geometry --- with the vacuum defined on some early times constant-khronon slice (on which static and freely follow observers basically coincide) --- then the evolution of the problem will uniquely determine the global vacuum state defined on this geometry.

\section{UV-completing the dispersion relation}
\label{sec:theral}
A word of caution must be shed here with respect to the derivation of Hawking radiation that we have performed in this work. Going back to our result, we found that the thermal spectrum of Hawking radiation emitted by a universal horizon was controlled by the peeling of rays close to it through
\begin{align}
    \bar{\kappa}=\frac{4}{3}\kappa_{\rm UH},
\end{align}
where $\kappa_{UH}=\frac{1}{2\mu}$ is the surface gravity of the universal horizon. The numerical coefficient in this result can be traced back to the behavior of the blue-shifted modes of the field $\phi$ departing close to the universal horizon, whose trajectory obeys the differential equation
\begin{align}
    \frac{\di t}{\di r}=\zeta_+=\frac{3}{2}\frac{\mu}{r-\mu}+\frac{3\Omega}{4\Lambda}\sqrt{ \frac{\mu}{r-\mu}}.
\end{align}

However, one must also note that the derivation in this paper was done perturbatively in large $\Lambda$, by expanding the dispersion relation \eqref{eq:dispersion} to leading order
\begin{align}
    \omega=\pm \sqrt{|F(r)|} \sqrt{q^2 + \frac{2q^4}{\Lambda^2}}= \pm\sqrt{|F(r)|} \left(q+\frac{q^3}{\Lambda^2}\right)+{\cal O}\left(\frac{1}{\Lambda^4}\right),
\end{align}
where therefore $\Lambda$ takes the role of an UV cut-off for our computation.

Immediately, a question thus arises, since the momenta of the out-going modes, ultimately responsible for the production of Hawking radiation, become of the order of the cut-off close to the universal horizon.
Indeed,
\begin{align}
    q^{\Lambda}_{\pm}= \Lambda\left(\pm\sqrt{\frac{\mu}{r-\mu}}+{\cal O}\left(\frac{1}{\Lambda}\right)\right).
\end{align}
So, although the perturbative expansion of the mode is under control, with sub-leading terms decaying appropriately with inverse powers of the cut-off and not hitting strong coupling at any point close to the universal horizon (where the expansion is valid), it is reasonable to question if we could be actually out of the validity regime of our perturbative approximation. Indeed, if instead of cutting the expansion of the dispersion relation at $q^3$ we retain the next suppressed correction
\begin{align}
    \omega =\pm \sqrt{|F(r)|}\left(q+\frac{q^3}{\Lambda^2}-\frac{q^5}{\Lambda^4}\right)+{\cal O}\left(\frac{1}{\Lambda^6}\right),
\end{align}
we do find that our result for $\bar\kappa$ \emph{changes} and becomes
\begin{align}
    \bar{\kappa}=\frac{8}{5}\kappa_{\text{\sc uh}},
\end{align}
with a different numerical pre-factor. Actually, if we cut the series at an order $\frac{q^N}{\Lambda^{N-1}}$ we find
\begin{align}\label{eq:kappa_N}
    \bar{\kappa}=\frac{2(N-1)}{N}\kappa_{\text{\sc uh}}.
\end{align}

This agrees with the results obtained in \cite{Ding:2016srk, Berglund:2012fk, Cropp:2016gkn}\footnote{Note however that there is a typo in \cite{Cropp:2016gkn}.} using other methods and confirms our suspicion that our result for the thermal spectrum actually lays outside the range of validity of the perturbative expansion on $\Lambda$, and it is actually controlled by the higher order polynomial term in the expansion for the momentum. However, there is hope. 

Instead of expanding the dispersion relation, and once we have understood how the mechanism of production of Hawking radiation around a universal horizon works, let us go back to the full dispersion relation \eqref{eq:dispersion}. Getting an analytical result with the full formula is hard because it requires to solve quartic equations that quickly become cumbersome and intractable. However, we do know now that Hawking radiation is solely controlled by the high energy behavior of the dispersion relation. We thus take, instead of the previous perturbative expansion, the opposite limit $\Lambda\rightarrow 0$, for which
\begin{align}
    \omega =\pm \sqrt{2|F(r)|}\ \frac{q^2}{\Lambda}+{\cal O}\left({\Lambda}^0\right),
\end{align}
thus focusing on the high-energy limit of the dispersion relation. Derivation of $\bar{\kappa}$ is now immediate from \eqref{eq:kappa_N} and leads to\footnote{We have also checked that this is the case by solving the equations exactly with the help of a computer.}
\begin{align}
    \bar{\kappa}=\kappa_{\text{\sc uh}},
    \label{eq:KUH}
\end{align}
which is the result that one would have actually expected from the beginning if the whole computation that we have performed here was never attempted!

Let us thus collect what we have learned here. If we were dealing with some unknown theory with dispersion relation given in a polynomial form
\begin{align}
    \omega=q+\dots +\frac{q^N}{\Lambda^{N-1}},
\end{align}
then the derivation that we have performed along this work applies exactly and one concludes that universal horizons emit Hawking Radiation with a spectrum controlled by $\bar \kappa$ as given by \eqref{eq:kappa_N}. However, if instead, we are dealing with a dispersion relation which is an analytic function -- but not a polynomial -- we have to focus on the limit of large $q$ of such dispersion relation, controlling the high energy behavior of the quantum modes in the theory, and again read the value of $\bar{\kappa}$ from our result~\eqref{eq:kappa_N}. Using this we find that for the case of the field with an action quartic in space derivatives (i.e.~$N=2$), $\bar{\kappa}$ exactly corresponds to the surface gravity of the universal horizon, eq.~\eqref{eq:KUH}.

The fact that Hawking's temperature depends on the specific form of the dispersion relation for different matter fields opens up an intriguing possibility, as discussed in detail in \cite{Dubovsky:2006vk}. Imagine that we have two different field species, that we denote as $\phi_1$ and $\phi_2$, whose dispersion relation at very high energies takes the form
\begin{align}
    &\omega_1^2\sim \frac{q^{N_1}}{\Lambda^{N_1-1}},\\
    &\omega_2^2\sim \frac{q^{N_2}}{\Lambda^{N_2-1}},
\end{align}
with $N_2>N_1$. Then, the corresponding Hawking radiation emitted by a black hole of the kind that we have studied in quanta of the fields $\phi_1$ and $\phi_2$ will have different temperature
\begin{align}
    &T_1=\frac{2(N_1-1)}{N_1}\frac{\kappa_{\rm UH}}{2\pi},\\
    &T_2=\frac{2(N_2-1)}{N_2}\frac{\kappa_{\rm UH}}{2\pi},
\end{align}
which implies $T_2>T_1$.

Now let us add two thin shells surrounding statically the black hole\footnote{Also known as Dyson spheres \cite{dyson_s}.} with temperatures $T_A$ and $T_B$ satisfying
\begin{align}
    T_2>T_B>T_A>T_1.
\end{align}

The key property of these shells will be that the shell $B$ only interacts with the matter particles $\phi_2$, while the shell $A$ only does so with particles of $\phi_1$. Then, this implies that the system will try to move towards thermodynamic equilibrium by radiating from $A$ onto the black hole using quanta of $\phi_1$ -- since $T_A>T_1$ -- and, at the same time, from the black hole onto $B$ using quanta of $\phi_2$. For an observer sitting at infinity and observing the system, the net effect of this heat transfer will be a flux from $A$ to $B$ mediated by the black hole which, since $T_B>T_A$, violates the second law of thermodynamics and allows for the construction of a mobile of the second kind.

Violation of the second law of thermodynamics is, of course, an undesirable property which in our case seems to hint towards the need for a universal dispersion relation for all matter species coupled to Einstein-Aether gravity. Indeed, if in the previous case $N_1=N_2$, then all kinds of radiation have the same Hawking temperature and the violation of the second law becomes impossible. This is what happens, for instance, in Ho\v rava Gravity, where all matter fields inherit the same universal momentum dependence on their dispersion relation due to matter-gravity interactions so that
\begin{align}
    \omega^2=c_2 q^2+\dots +c_{2d}\frac{q^{2d}}{\Lambda^{2d-2}},
\end{align}
where the coefficients $c_{i}$ are species-dependent and $d$ is the number of spatial dimensions, thus $d=3$ for a $3+1$ dimensional space-time.
This implies that generically we should expect a universal spectrum with $\bar{\kappa}=4/3 \kappa_{\text{\sc uh}}$ (species-dependent constants like $c_{2d}$ being always removable by a simple redefinition of the  cut-off scale $\Lambda$ which does not appear in the final expression for the temperature).

%%%%%%%%%%%%%%%%%%%%%%%%%%%%%%%%%%%%%%%%%%%%%
\section{Discussion and conclusions}

In summary, we have shown in two alternative ways that the Hawking radiation characterizing a black hole endowed with a universal horizon is set by the effective surface gravity characterizing the peeling of physical null rays in its proximity Eq.~\eqref{eq:kappa_N}. This result implies that the temperature generically is not just set by the surface gravity of the universal horizon Eq.~\eqref{eq:kappaUH} but depends on the high energy behaviour of the dispersion relation of the specific field considered. Robustness of the black hole thermodynamics requires then that such UV behaviour of the dispersion relation should be universal (to avoid different temperatures of different fields on the same black hole space-time), like for example the one expected in Ho\v rava gravity.
While we deem these results a definite progress in understanding the internal consistency of these theories and the basic building blocks of black hole thermodynamics, there are nonetheless some open issues that still need to be addressed. 

Although we have explicitly shown that Hawking temperature is determined by the UV behaviour of the field dispersion relation, one has to be aware that the black hole solutions with universal horizon have been so far derived only as solutions of the low energy gravitational Lagrangian (for example in Ho\v rava gravity the so called $L_2$ part, providing just second order field equations for the gravitational field). It is not known if the UV completion of the gravitational Lagrangian will still admit such black holes as solutions, nor if the modified solution will still admit universal horizons. However, an examination of curvature invariants on the universal horizon, which can be arbitrarily small, seems to indicate that one can always find a situation in which the UV-completing terms -- given in terms of higher powers of curvature invariants, at least in Ho\v rava Gravity -- can be always taken as small perturbations. If this is the case, it seems plausible that the universal horizon will survive in the full theory with similar properties to the ones described here. It is true nonetheless that a full account of its dynamics requires the knowledge of the UV-complete theory up to arbitrary energy. 

Another point worth further investigating is related to the difference between our computation based on the WKB approximation and the one reported in~\cite{Michel:2015rsa}. As we have seen, the two derivations lead to different conclusions just because of a different choice of the vacuum state. In our derivation, we assumed that the regular state on the black hole space-time is the analogue of the standard Unruh state for physical free falling observers (which do not follow the geodesics of the metric but rather those determined with the metric-aether background). In the case of~\cite{Michel:2015rsa} a vacuum state defined with respect to the aether frame was chosen instead. Which of the two choices is the physically relevant one, cannot be settled just by the concordance of the second derivation with the first one, based on matching regular sets of coordinates along a physical ray trajectory, as the latter implicitly assumes the regularity of the vacuum state all along these worldlines. 

In relativistic derivations, it can be shown exactly that the Unruh state 
%(which appears as vacuum at past null infinity and at horizon crossing for free falling observers while being thermal at future null infinity at late times), 
is the only regular state compatible at late times with a collapsing geometry leading to the formation of a horizon (see e.g.~\cite{Kay:1988mu,Barcelo:2007yk})\footnote{Note that the vacuum state does not evolve and is globally defined on the whole space-time, as standard QFT is defined within the Heisenberg representation.}. 
However, some ambiguity arises in non-relativistic theories as no infinite accumulation of physical rays happens anymore at the Killing horizon (due to the modified dispersion relation) and hence no infinities are expected there in the energy distributions in any frame. Considering non-relativistic theories of gravity as well, led to the discovery of universal horizons which may remove such ambiguity by restoring the peeling behaviour typical of relativistic theories and thus reintroducing the key element for the singular behaviour of global quantum states which are not vacuum at (universal) horizon crossing. Nonetheless, an ambiguity remains due to the fact that such regularity can be defined w.r.t. alternative frames: the one for physical freely falling observers and the one for observers falling with the aether frame. In this work, we have deemed the first one as the physically relevant frame, but only an eventual calculation of the renormalized stress energy tensor analogue to the one carried out in relativistic physics will be able to address the uniqueness and regularity of the vacuum state on a collapsing geometry in a non-relativistic framework.  

In order to do this, several road blocks will have to be removed, for example, the study of the Hadamard condition for Green functions (see e.g.~\cite{birrell1984quantum}) in non-relativistic settings, as well as the generalization to space-times with universal horizons of the theorems concerning vacuum state regularity in globally hyperbolic space-times \cite{Fulling:1978ht,Fulling:1981cf,Kay:1988mu}. Let us stress that such generalizations are far from clear in the quite contrived causality structure of such geometries~\cite{Carballo-Rubio:2020ttr} and that the possibility to fully characterize the vacuum state is not guaranteed due to the peculiar nature of universal horizons that make them similar to Cauchy horizon in non-relativistic frameworks~\cite{Blas:2011ni, Carballo-Rubio:2020ttr}.~\footnote{
One way to understand how a universal horizon could act as a Cauchy horizon in non-relativistic settings is to realize that leaves of the preferred foliation before the formation of the universal horizon all asymptote to $i^0$ (see figure~\ref{fig:penrose}), and so to get a well-defined Cauchy problem at worst one needs to impose some regularity condition at spacelike infinity $i^0$. However, after the formation of a universal horizon, in order to set up a well-defined Cauchy problem, one needs new extra ``initial data" (corresponding to some regularity condition at future timelike infinity $i^+$). See e.g.~\cite{Carballo-Rubio:2020ttr} for a more detailed discussion.}
We hope to be able to tackle these issues and more in forthcoming investigations.

\section*{Acknowledgements}
We are grateful to Enrico Barausse, Ted Jacobson, David Mattingly and Sergey Sibiryakov for discussions. M. H-V. has been supported by the European Union's H2020 ERC Consolidator Grant “GRavity from Astrophysical to Microscopic Scales” grant agreement no. GRAMS-815673. S.L. acknowledge funding from the Italian Ministry of Education and Scientific Research (MIUR) under the grant PRIN MIUR 2017-MB8AEZ. R. S-G. has been supported by the Spanish FPU Grant No FPU16/01595 and by COST action CA16104 ``GWverse"  through a Short Term Scientific Mission. R. S-G. also wishes to thank the Astroparticle Physics group at SISSA for their hospitality during early stages of this work. 

%%%%%%%%%%%%%%%%%%%%%%%%%%%%%%%%
%%%%%%%%%%%%%%%%%%%%%%%%%%%%%%%%
%%%%%%%%%%%%%%%%%%%%%%%%%%%%%%%%
%%%%%%%%%%%%%%%%%%%%%%%%%%%%%%%%

\appendix
%%%%%%%%%%%%%%%%%%%%%%%%%%%%%%%%
%%%%%%%%%%%%%%%%%%%%%%%%%%%%%%%%
%%%%%%%%%%%%%%%%%%%%%%%%%%%%%%%%
%%%%%%%%%%%%%%%%%%%%%%%%%%%%%%%%

%%%%%%%%%%%%%%%%%%%%%%%%%%%%%%%%
%%%%%%%%%%%%%%%%%%%%%%%%%%%%%%%%
%%%%%%%%%%%%%%%%%%%%%%%%%%%%%%%%
%%%%%%%%%%%%%%%%%%%%%%%%%%%%%%%%
\bibliography{hawking}{}
\end{document}